\documentclass[10pt,aps,prb,twocolumn,superscriptaddress]{revtex4-2}
\usepackage{amsmath}
\usepackage{fixltx2e}
\usepackage{hyperref}
\usepackage{graphicx}
\usepackage{subfigure}
\usepackage{xspace}

\usepackage[dvipsnames]{xcolor}

\usepackage{amsfonts}
\usepackage{mathrsfs}
\usepackage{dsfont}
\usepackage{mathtools}
\usepackage[capitalize]{cleveref}
\usepackage{simplewick}
\usepackage{hyperref}
\usepackage{float}
\usepackage[normalem]{ulem}

\usepackage{tabularx}
\usepackage{amstext} %
\usepackage{array}   %
\usepackage{multirow}

\newcolumntype{L}{>{$}l<{$}} %
\newcolumntype{C}{>{$}c<{$}} %

\begin{document}

\newcommand\kr[1]{\textcolor{blue}{(KR: #1)}}
\newcommand\dk[1]{\textcolor{magenta}{(DK: #1)}}
\newcommand\rv[1]{\textcolor{orange}{(RV: #1)}}
\renewcommand\sb[1]{\textcolor{red}{(SB: #1)}}
\newcommand\ar[1]{\textcolor{brown}{(AR: #1)}}
\newcommand{\corout}[1]{{\color{black}\sout{#1}}}

\title{Pressure study on the interplay between magnetic order and valence-change crossover in EuPd$_2$(Si$_{1-x}$Ge$_x$)$_2$}

\author{Bernd Wolf}
\affiliation{Physikalisches Institut, Goethe-Universit\"at, 60438 Frankfurt (M), Germany}
\author{Theresa Lundbeck}
\affiliation{Physikalisches Institut, Goethe-Universit\"at, 60438 Frankfurt (M), Germany}
\author{Jan Zimmermann}
\affiliation{Physikalisches Institut, Goethe-Universit\"at, 60438 Frankfurt (M), Germany}
\author{Marius Peters}
\affiliation{Physikalisches Institut, Goethe-Universit\"at, 60438 Frankfurt (M), Germany}
\author{Kristin Kliemt}
\affiliation{Physikalisches Institut, Goethe-Universit\"at, 60438 Frankfurt (M), Germany}
\author{Cornelius Krellner}
\affiliation{Physikalisches Institut, Goethe-Universit\"at, 60438 Frankfurt (M), Germany}
\author{Michael Lang}
\affiliation{Physikalisches Institut, Goethe-Universit\"at, 60438 Frankfurt (M), Germany}


\date{\today}

\begin{abstract}
We present results of the magnetic susceptibility on high-quality single crystals of EuPd$_2$(Si$_{1-x}$Ge$_x$)$_2$ for Ge concentrations 0 $\leq x \leq$ 0.105 performed under varying hydrostatic (He-gas) pressure 0 $\leq p \leq$ 0.5\,GPa. The work extends on recent studies at ambient pressure demonstrating the drastic change in the magnetic response from valence-change-crossover behavior for $x$ = 0 and 0.058, to long-range antiferromagnetic (afm) order below $T_{\text{N}}$ = 47\,K for $x$ = 0.105. The valence-change-crossover temperature $T'_{\text{V}}$ shows an extraordinarily strong pressure dependence of d$T'_{\text{V}}$/d$p$ = +(80 $\pm$ 10)\,K/GPa. In contrast, a very small pressure dependence of d$T_{\text{N}}$/d$p \leq$ +(1 $\pm$ 0.5)\,K/GPa is found for the afm order upon pressurizing the $x$ = 0.105 crystal from $p$ = 0 to 0.05\,GPa. Remarkably, by further increasing the pressure to 0.1\,GPa, a drastic change in the ground state from afm order to valence-change-crossover behavior is observed. Estimates of the electronic entropy related to the Eu 4$f$ electrons, derived from analyzing susceptibility data at varying pressures, indicate that the boundary between afm order and valence-change crossover represents a first-order phase transition. Our results suggest a particular type of second-order critical endpoint of the first-order transition for $x$ = 0.105 at $p_{\text{cr}} \approx$ 0.06\,GPa and $T_{\text{cr}} \approx$ 39\,K where intriguing strong-coupling effects between fluctuating charge-, spin- and lattice degrees of freedom can be expected.
\end{abstract}


\maketitle

\section{Introduction}
\subsection{Valence transition and critical endpoint}

In the vicinity of a second-order critical endpoint (CEP) which terminates a first-order phase transition line, strong fluctuations are expected which provide deep insight into the system's universal properties. Of particular interest are materials with correlated electrons tuned to a CEP by the application of pressure. For these systems strong-coupling effects between the correlated electrons and the lattice degrees of freedom can be expected \cite{Zacharias2015QCElasticity}. As an example we mention the phenomenon of \textit{critical elasticity} recently observed upon pressure tuning an organic Mott insulator close to the CEP of the first-order Mott transition line \cite{Gati2016breakdown}. A pronounced softening of the lattice was observed over a considerably wide $T-p$ range around the CEP, indicating a particular strong coupling between the critical electronic system and the lattice degrees of freedom.\\

Similar strong-coupling effects can be expected also for other CEPs amenable to pressure tuning. In this regard, the valence-transition CEP represents a particularly interesting scenario as this transition involves, besides the charge- and lattice degrees of freedom, also significant changes in the system's magnetic properties. Rare earth-based intermetallics have been intensively used as target materials for studying valence fluctuations and their interplay with magnetism \cite{Lawrence1981valence, Ryan2023VTMag, Hossain2004VT, Ahmida2020VF, Baenitz2006NMR, Dionicio2006VT, Hossain2004AFM, Muthu2016phase, Nakamura2015trans, Gouchi2020AFM, Fukuda2003}. Based on these investigations a temperature-pressure ($\textit{T-p}$) phase diagram for serveral Eu-based compounds has been derived \cite{Onuki2017review,Wada1999phase,Onuki2020phase,Cho2002poly,Seiro2011EuIr2Si2}. At low pressures, the Eu ions are in their large-volume Eu$^{2+}$ states which carry a local magnetic moment. The coupling between these moments, mediated via the Ruderman-Kittel-Kasuya-Yoshida (RKKY) interaction, usually gives rise to long-range antiferromagnetic order at $T \leq T_{\text{N}}$. By the application of pressure the system adopts a non-magnetic low-volume Eu$^{(3-\delta)+}$ ($0 < \delta <$ 1) state by crossing a first-order valence-transition line $T_{\text{V}}(p)$ (for $T > T_{\text{N}}$) \cite{Onuki2017review,Batlogg1982EuPd2Si2}. This first-order line $T_{\text{V}}(p)$ terminates at a second-order CEP. On the high-pressure side of the CEP crossover behavior is expected and experimentally observed \cite{Onuki2017review,Onuki2020phase,Seiro2011EuIr2Si2}.

\subsection{The case of EuPd$_2$Si$_2$ }

There are a few cases where a valence change can be induced by varying the temperature at ambient pressure \cite{Batlogg1982EuPd2Si2,Mimura2004EuPd2Si2,Felner1986YbInCu4}. Among them is EuPd$_2$Si$_2$ \cite{Sampath1981EuPd2Si2}, which crystallizes in the tetragonal ThCr$_2$Si$_2$ structure. Upon cooling, the system shows a pronounced but still continuous valence change from Eu$^{2.3+}$ at high temperatures (around 300\,K) to Eu$^{2.8+}$ below about 100\,K \cite{Sampath1981EuPd2Si2, Croft1982config, Mimura2004EuPd2Si2}. This valence-change crossover manifests itself in a slightly broadened drop in the magnetic susceptibility within a narrow temperature interval of about 40\,K \cite{Onuki2017review, Kliemt2022cryst, Peters2022cryst,Wolf2022SCES}. A crossover temperature $T'_{\text{V}} \approx$ 160\,K has been assigned by using the position where the change in the magnetic susceptibility is largest \cite{Peters2022cryst,Wolf2022SCES}. The notion of a valence-change crossover in EuPd$_2$Si$_2$ is consistent with earlier findings on polycrystalline material \cite{Segre1982mix,Batlogg1982EuPd2Si2}, locating the system on the high-pressure side of the CEP. The research on this material has regained momentum recently thanks to the success in growing large single crystals of pure EuPd$_2$Si$_2$ \cite{Onuki2017review,Kliemt2022cryst} and Ge-substituted EuPd$_2$(Si$_{1-x}$Ge$_x$)$_2$ \cite{Peters2022cryst}.\\

In the present work we focus on the effects of Ge-substitution and hydrostatic pressure on the valence-change-crossover behavior, aiming at identifying suitable parameters by which the system can be tuned close to its CEP. Details on the single crystal growth and sample characterization via magnetic susceptibility, thermal expansion and structural investigations can be found in Refs.\,\cite{Kliemt2022cryst, Peters2022cryst}, see also Ref.\,\cite{Wolf2022SCES} for a preliminary account of magnetic susceptibility and thermal expansion results as well as Ref.\,\cite{Ye2022Raman} for Raman-spectroscopy data. By employing susceptibility measurements in combination with fine (He-gas) pressure tuning under truly hydrostatic pressure conditions, we find strong indications for the existence of a particular type of CEP, referred to as "CEP", for single crystalline EuPd$_2$(Si$_{1-x}$Ge$_x$)$_2$ with $x$ = 0.105. We argue that, different from the valence-transition CEP implied in the generalized phase diagram \cite{Onuki2017review,Wada1999phase,Onuki2020phase,Cho2002poly,Seiro2011EuIr2Si2}, where the valence-crossover state emerges from a paramagnetic phase, the "CEP" for $x$ = 0.105 marks the endpoint of a first-order transition line separating long-range afm order from a valence-change-crossover state. Our results indicate that the "CEP" for the $x$ = 0.105 compound is located at $p_{\text{cr}} \approx$ 0.06\,GPa and $T_{\text{cr}} \approx$ 39\,K making it readily accessible for future experiments aiming at a detailed investigation of the expected strong-coupling effects.

\section{Methods}
\subsection{Experiments}
Single crystals of EuPd$_2$(Si$_{1-x}$Ge$_x$)$_2$ with Ge-concentrations $x$ = 0  (\#1, \#2), 0.058 (\#3), and 0.105 (\#4, \#5) were grown by using the Czochralski method. Table 1 gives a list of the crystals investigated along with some characteristic temperatures and their pressure dependence as obtained in the present study. Details of the crystal growth, the determination of the real germanium concentration by energy dispersive X-ray analysis (EDX), and results of some basic structural-, magnetic-, and thermodynamic investigations can be found in Refs.\,\cite{Kliemt2022cryst, Peters2022cryst, Wolf2022SCES}. In what follows we refer to these crystals by their real Ge-concentration and specify each individual sample by a sample number. The susceptibility was measured by using a commercial superconducting quantum interference device (SQUID) magnetometer (MPMS, Quantum Design) equipped with a CuBe pressure cell (Unipress Equipment Division, Institute of High Pressure Physics, Polish Academy of Science). The pressure cell is connected via a CuBe capillary to a room-temperature He-gas compressor, serving as a gas reservoir. This setup enables temperature sweeps over wide temperature ranges 2\,K $\leq T \leq$ 300\,K to be performed at almost ideal $p$ = const. conditions, up to a maximum pressure of $p_{\text{max}}$ = 0.6\,GPa, see Ref.\,\cite{Wolf2022RuCl3} for details. For the interpretation of the experimental results, it is important to note that for all pressure experiments reported here, helium is in its liquid state ensuring ideal hydrostatic-pressure conditions.\\

\begin{table}
\centering
\caption[]{List of EuPd$_2$(Si$_{1-x}$Ge$_x$)$_2$ single crystals investigated. Given are the real Ge-concentrations $x_{\text{real}}$ determined by EDX analysis \cite{Peters2022cryst}, the individual sample number, the valence-crossover temperature $T'_{\text{V}}$ and its pressure dependence d$T'_{\text{V}}$/d$p$ as well as the N\'{e}el temperature $T_{\text{N}}$ characterizing the transition into antiferromagnetic order.  }
\label{tab:1}

%
\renewcommand{\arraystretch}{1.5}
\setlength\tabcolsep{5pt}
\begin{tabular}{@{}ccccc@{}}
\hline\noalign{\smallskip}
\textbf{$x_{real}$} & \text{sample} & $T'_{\text{V}}$  & d$T'_{\text{V}}$/d$p$   & $T_{\text{N}}$     \\
& \text{no.} &  [K] & [K/GPa] &  [K]   \\
\hline\noalign{\smallskip}
             &   & $p = 0$  &   & $p = 0$ \\
\hline\noalign{\smallskip}
0 & \text{\#1} & 155 $\pm$ 1.5   & 80 $\pm$ 5   & - \\
0  & \text{\#2} &  120 $\pm$ 1.5  & 80 $\pm$ 5   & - \\
0.058(7)  & \text{\#3} & 90 $\pm$ 1.5  & 75 $\pm$ 5  & - \\
\hline\noalign{\smallskip}
             &   & $p = 0.1$ GPa & $p \geq 0.1$ GPa & $p = 0$ \\
\hline\noalign{\smallskip}
0.105(8)  & \text{\#4} & 45 $\pm$ 1.3  & 90 $\pm$ 10 & 47.3 $\pm$ 0.2   \\
0.105(8) & \text{\#5} & 45 $\pm$ 1.3  &  100 $\pm$ 10 &  47.3 $\pm$ 0.2  \\
\hline
\end{tabular}
\end{table}

\section{Experimental results}
\subsection{Overview of the magnetic behavior for crystals with 0 $\leq x \leq$ 0.105 at ambient pressure}

\begin{figure}[h]
	\includegraphics[width=1.0\columnwidth]{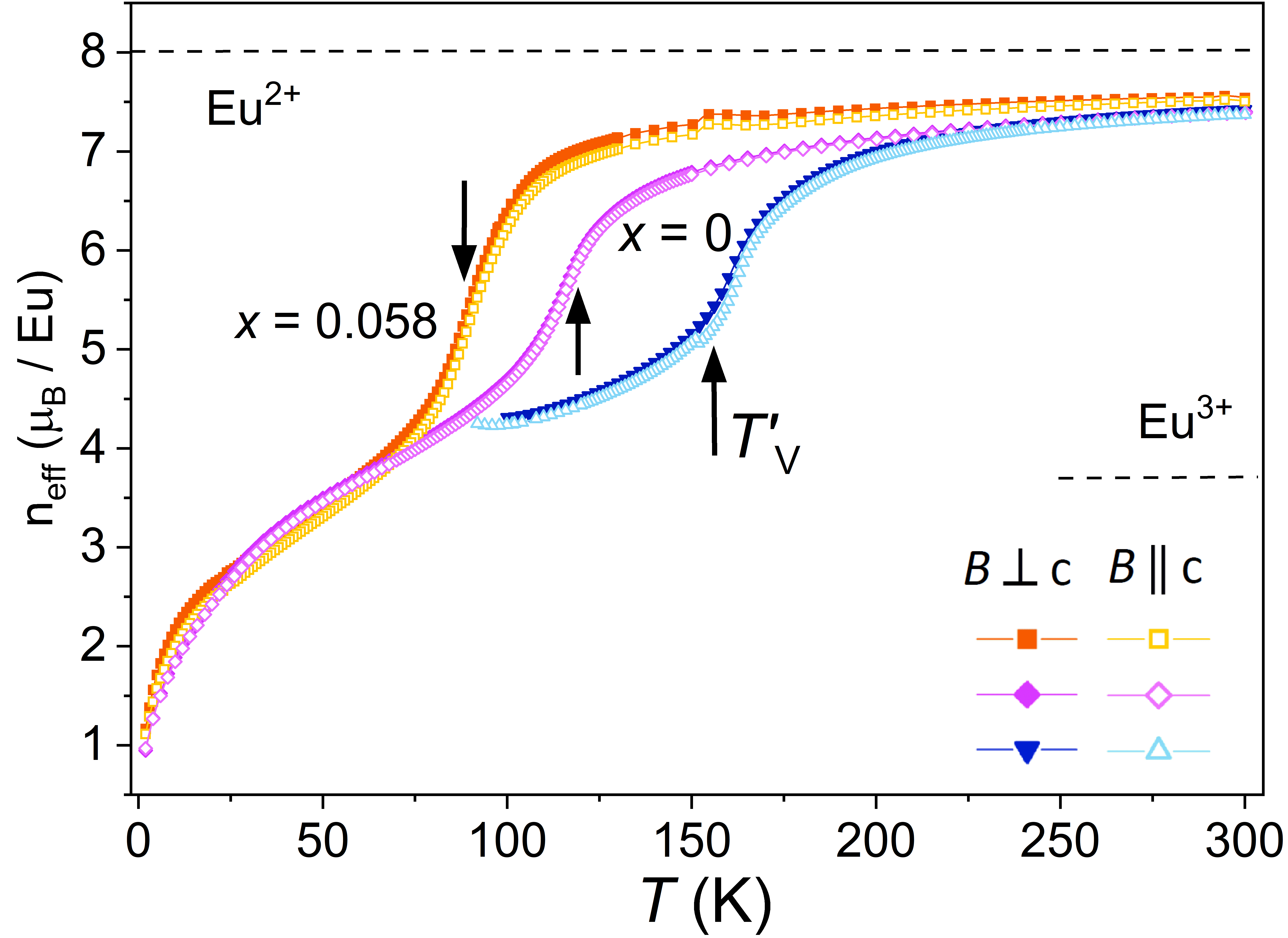}
	\caption{Effective magnetic moment as a function of temperature for single crystals of EuPd$_2$(Si$_{1-x}$Ge$_x$)$_2$ with $x$ = 0 [\#1 (blue symbols) and \#2 (purple symbols)] and $x$ = 0.058 (\#3). The data were taken in a magnetic field of 1\,T for orientations parallel (open symbols) and perpendicular (closed symbols) to the tetragonal $c$ axis. The dashed lines represent the full effective moment of free Eu$^{2+}$ (4$f^{7}$) and Eu$^{3+}$ (4$f^{6}$) ions at 300\,K with a typical energy difference of $\sim$ 450\,K between the ground state and the first excited state \cite{Takikawa2010}.
Arrows indicate the valence-crossover temperature $T'_{\text{V}}$ derived from the maximum position in d$(\chi \cdot T$)/d$T$. The small anomalies around 150\,K are of extrinsic nature.}
	\label{Fig1}
\end{figure}

 In FIGs.\,1 and 2 we give an overview of the different magnetic behaviors revealed for single crystalline EuPd$_2$(Si$_{1-x}$Ge$_x$)$_2$ in response to small changes in the Ge content from $x \leq$ 0.058 (FIG.\,1) to $x$ = 0.105 (FIG.\,2). To this end we plot the effective magnetic moment $n_{\text{eff}}$ and its evolution with temperature for 2\,K $\leq T \leq$ 300\,K as derived from magnetic susceptibility data $\chi$($T$) \cite{Peters2022cryst, Wolf2022SCES} via $n_{\text{eff}}$ = 2.828 $\cdot$ [$\chi$($T$)$\cdot T$]$^{1/2}$ in cgs units by using a spectroscopic $g$ factor of 2. By plotting the data in this representation, small $T$-induced changes in the magnetic moment at high temperatures can be easily discerned. The measurements were carried out at ambient pressure, labelled $p$ = 0 from here on, by applying a magnetic field of $B$ = 1\,T both parallel (open symbols) and perpendicular (closed symbols) to the tetragonal $c$ axis. Figure 1 shows $n_{\textrm{eff}}$($T$) of two non-substituted single crystals $x = 0$ (\#1, \#2) and of a crystal with $x = 0.058$ (\#3).
 The data for $x$ = 0 show a gradual reduction of $n_{\text{eff}}$ on cooling from 300\,K, followed by a more rapid drop within a rather narrow temperature window around 155\,K (120\,K) for crystal \#1 (\#2). The significant reduction in $n_{\text{eff}}$ to a value close to that of Eu$^{3+}$ is assigned to the temperature-induced valence-change crossover from Eu$^{(2+\delta)+}$ to Eu$^{(3-\delta')+}$ \cite{Peters2022cryst, Wolf2022SCES}, consistent with earlier results on polycrystalline material for $x$ = 0 \cite{Segre1982mix, Wada2001mix}. To parameterize this crossover behavior, we use the quantity d$(\chi \cdot T$)/d$T$ (cf. inset of FIG.\,3 and FIG. 11 in Appendix C for details), yielding Lorentzian-shaped curves within the crossover regime. We refer to the maximum in d$(\chi \cdot T$)/d$T$ as the valence-crossover temperature $T'_{\text{V}}$, see Table 1 for a compilation of the so-derived $T'_{\text{V}}$ values. Further characteristics of the data for the $x$ = 0 crystals in FIG.\,1 include a significant variation of $T'_{\text{V}}$ for the two $x$ = 0 crystals \#1 and \#2 investigated, and an almost isotropic magnetic behavior throughout the entire temperature range investigated, see also inset of FIG.\,2. For the Ge-substituted crystal with $x$ = 0.058 (\#3), we observe a similar behavior in $n_{\text{eff}}$($T$) albeit with a significant reduction of $T'_{\text{V}}$ down to 90\,K accompanied by an enhanced drop in $n_{\text{eff}}$ down to about the same value as observed for $x$ = 0. The data for $x \leq$ 0.058 shown in FIG.\,1 all share the same weak magnetic anisotropy, i.e., an $n_{\text{eff}}$($B\perp c$) which only slightly exceeds $n_{\text{eff}}$($B \parallel c$) throughout the entire temperature range investigated.


\begin{figure}[h]
	\includegraphics[width=1.0\columnwidth]{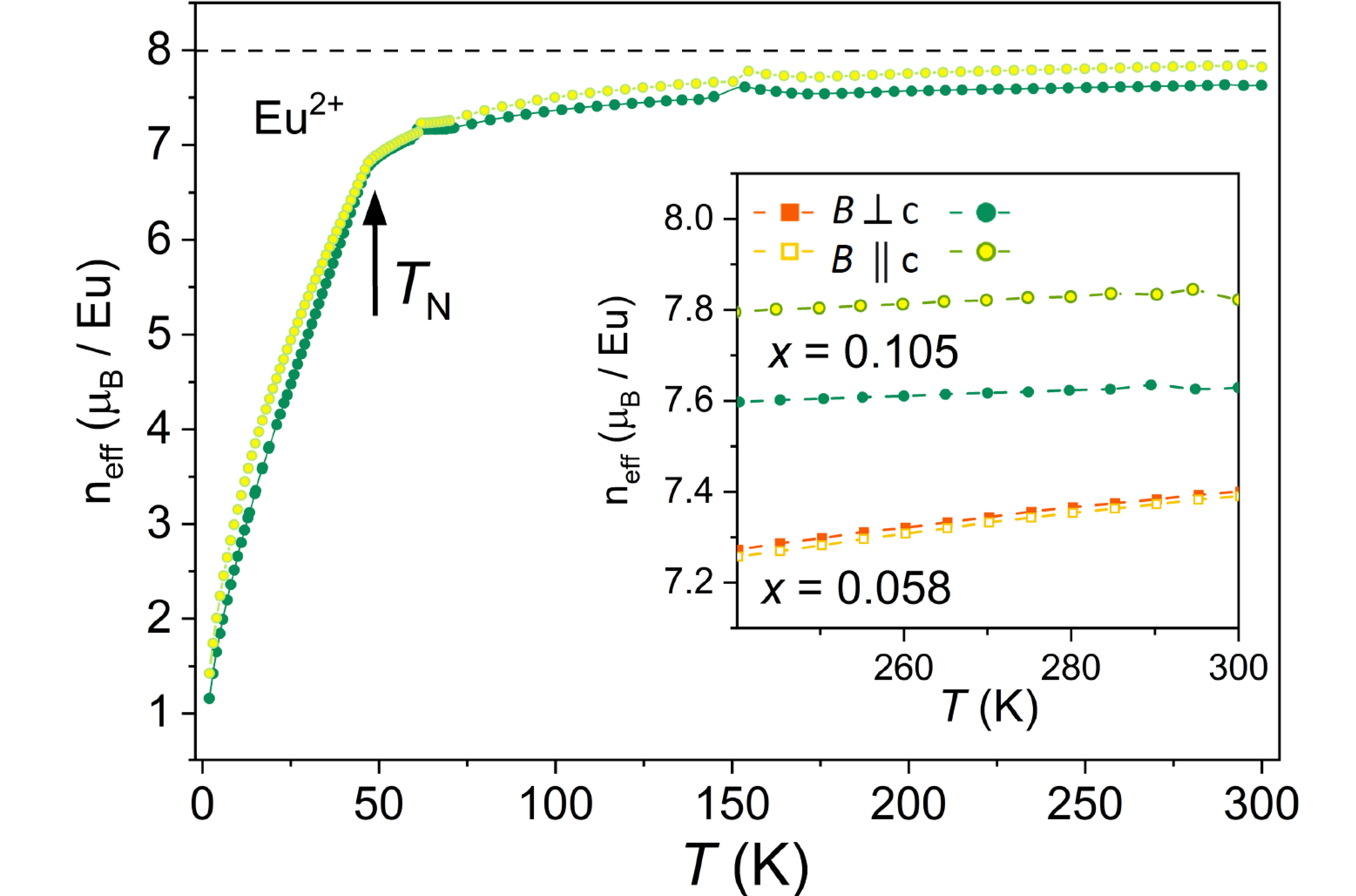}
	\caption{Effective magnetic moment $n_{\text{eff}}$ as a function of temperature for a single crystal of EuPd$_2$(Si$_{1-x}$Ge$_x$)$_2$ with $x$ = 0.105 (\#5). The data were taken in a magnetic field of 1\,T for orientations parallel (open symbols) and perpendicular (closed symbols) to the tetragonal $c$ axis. The dashed horizontal line represents the full effective moment of free Eu$^{2+}$ (4$f^{7}$) ions. The small anomalies around 150\,K and 75\,K are of extrinsic nature. The inset shows a blow-up of the data for $T >$ 240\,K for highlighting the magnetic anisotropy. For comparison the inset also includes the data for the crystal with $x$ = 0.058 (\#3)(cf. FIG.\,1) in the same temperature range.}
	\label{Fig1}
\end{figure}

 In FIG.\,2 we show $n_{\text{eff}}$($T$) for 2\,K $\leq T \leq$ 300\,K for a single crystal with $x$ = 0.105 (\#5). Almost identical data (not shown) were obtained for a second crystal (\#4) of the same substitution level with only small differences for temperatures below 20\,K. The data reveal a distinctly different behavior from that observed for $x \leq$ 0.058 in FIG.\,1, in showing a sharp kink at around 47\,K which separates a slowly varying and somewhat enhanced $n_{\text{eff}}$ at high temperatures from a rapidly decreasing $n_{\text{eff}}$ down to lowest temperatures. We assign this behavior to long-range afm order below $T_{\text{N}}$ = 47.3\,K, see Ref.\,\cite{Peters2022cryst, Wolf2022SCES}  for detailed investigations in support of this claim, including magnetic-, thermodynamic- and structural investigations. In particular the study in Ref.\,\cite{Peters2022cryst} revealed an easy-plane magnetic anisotropy below $T_{\text{N}}$ which is typically observed in Eu-based magnets, see also Sect. IV. C. \\

Another, more subtle difference in the magnetic behavior of the $x$ = 0.105 crystal as opposed to that observed for lower Ge content relates to the magnetic anisotropy in the paramagnetic regime $T > T_{\text{N}}$. As shown in the inset of FIG.\,2 we find that the anisotropy for the crystal with $x$ = 0.105 is enhanced and reversed, now yielding an $n_{\text{eff}}$($B \parallel c$) which is significantly larger than $n_{\text{eff}}$($B\perp c$). This observation indicates changes in the local Eu$^{2+}$ environment for the $x$ = 0.105 samples, exhibiting afm order, as opposed to the crystals showing valence-crossover behavior (see Discussion subsection B for more details).


\subsection{Magnetic susceptibility at varying hydrostatic pressure}

\begin{figure}[h]
	\includegraphics[width=1.0\columnwidth]{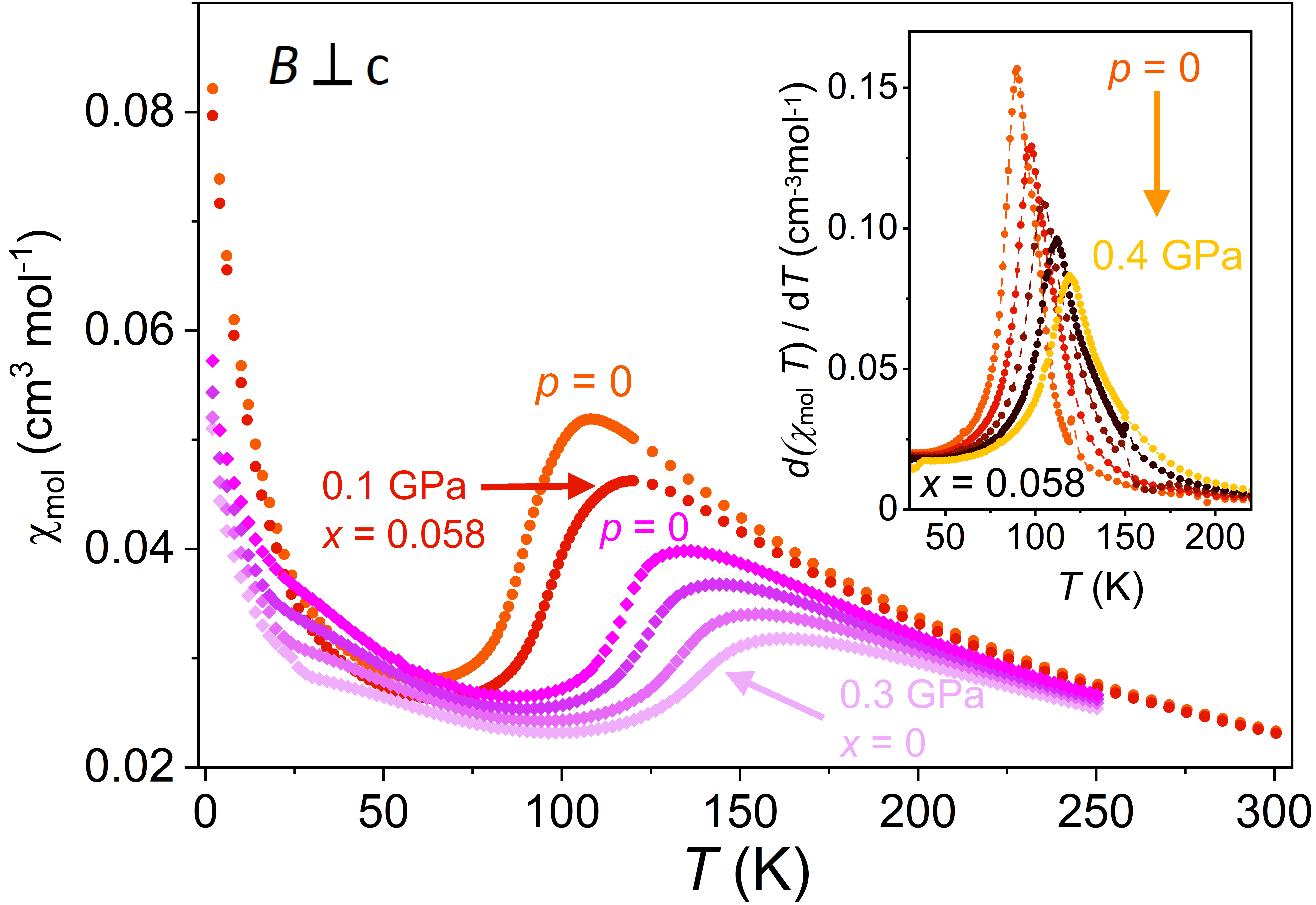}
	\caption{Molar magnetic susceptibility as a function of temperature of single crystals of EuPd$_2$(Si$_{1-x}$Ge$_x$)$_2$ with $x$ = 0 (\#2) and $x$ = 0.058 (\#3) at varying pressures $p \leq$ 0.4\,GPa. The magnetic field of $B$ = 1\,T was oriented perpendicular to the tetragonal $c$ axis. The inset exhibits the temperature derivative of $\chi \cdot T$ of the crystal with $x$ = 0.058 (\#3) for 0 $\leq p \leq$ 0.4\,GPa in a narrow temperature window around $T'_V$.}
	\label{Fig1}
\end{figure}

 After having recapitulated the marked changes revealed in the magnetic response from valence-change-crossover behavior for $x \leq$ 0.058 to long-range afm order for $x$ = 0.105, we will now focus on the effect of hydrostatic pressure on the respective behavior. To this end we show in FIG.\,3 susceptibility data for $B$ perpendicular to the $c$ axis, $\chi_{\perp}$, for crystals with $x$ = 0 (\#2) and $x$ = 0.058 (\#3) as a function of temperature at varying hydrostatic pressures $p \leq$ 0.4\,GPa. For both substitution levels we find pronounced changes in $\chi_{\perp}$ with pressure at intermediate temperatures, manifesting themselves in a significant shift of the valence-change-crossover temperature $T'_{\text{V}}$ to higher temperatures, accompanied by a pronounced broadening of the crossover region. This is shown in more detail in the inset of FIG.\,3, where the quantity d($\chi \cdot T)$/d$T$ for crystal \#3 ($x$ = 0.058) is plotted for varying pressures $0 \leq p \leq 0.4$\,GPa in a narrow temperature window around $T'_{\text{V}}$. By identifying the position of the maximum with $T'_{\text{V}}$, we find a pressure dependence of d$T'_{\text{V}}$/d$p$ = +(75 $\pm$ 5)\,K/GPa, cf.\,Table 1. These curves, which can be well described by a Lorentzian function (see FIG.\,11 in Appendix C), also enable us to quantify the width of the valence-change-crossover region by using the full width at half maximum $\Gamma$ of the Lorentzian. The crossover temperature $T'_{\text{V}}$ as well as the width $\Gamma$ will be used below (FIG.\,6 and FIG.\,7) for constructing the $T-p$ phase diagram. Figure 3 demonstrates that there is a strong effect of pressure on $\chi_{\perp}$ for temperatures within the valence-change-crossover region. In contrast, there is only a minor effect pressure has on $\chi_{\perp}$ outside this regime, i.e., for temperatures above about 220\,K and below about 25\,K. In particular, we mention the strong upturn in $\chi_{\perp}$ at low temperatures which is practically unaffected by the applied pressure. We consider this upturn as a characteristic feature of all crystals showing a valence crossover, see section IV.B. below. \\


\begin{figure}[h]
	\includegraphics[width=1.0\columnwidth]{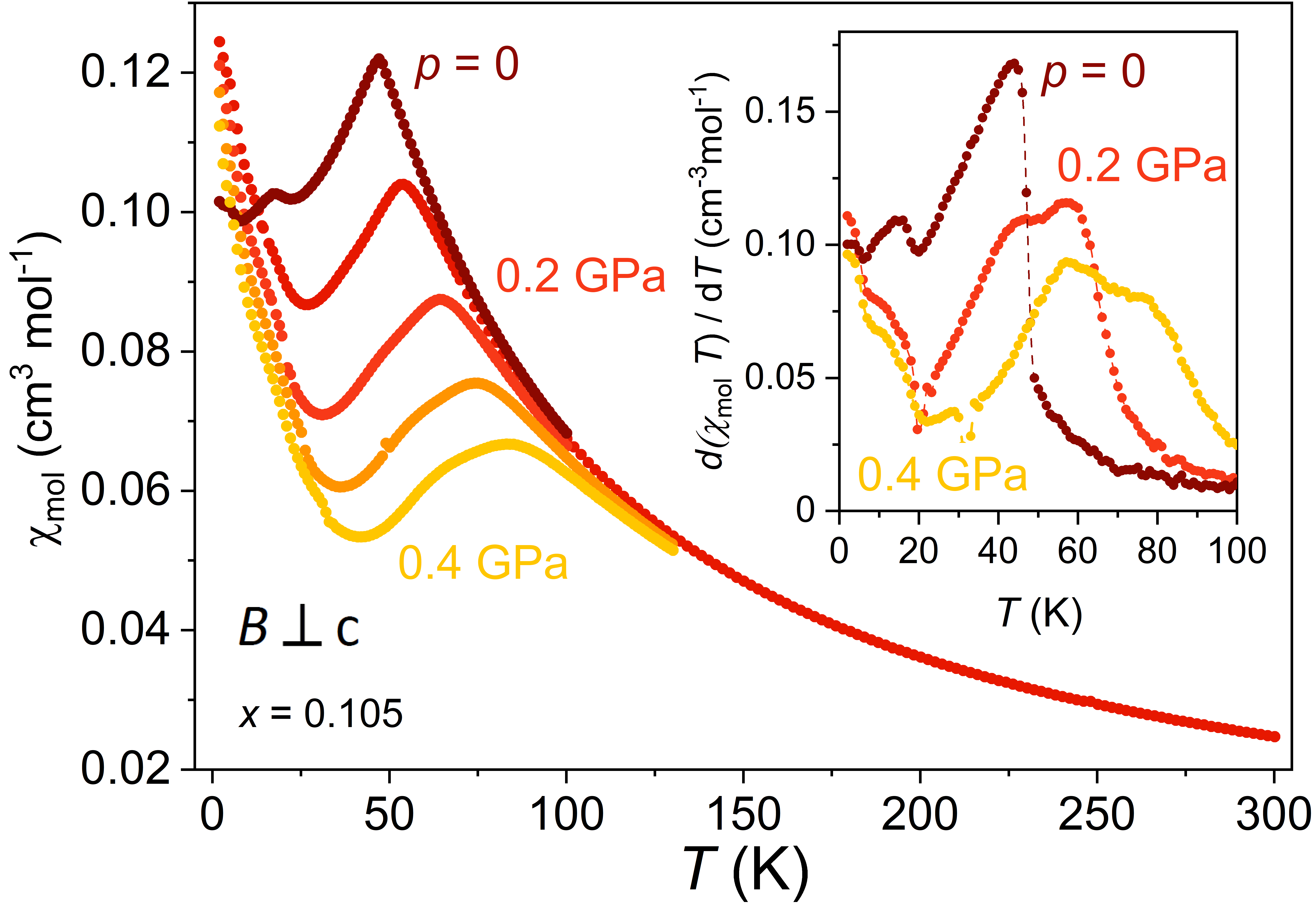}
	\caption{Molar magnetic susceptibility as a function of temperature of a single crystal of EuPd$_2$(Si$_{1-x}$Ge$_x$)$_2$ with $x$ = 0.105 (\#4) for varying pressure 0 $\leq p \leq$ 0.4\,GPa. The external magnetic field of $B$ = 1\,T was oriented perpendicular to the tetragonal $c$ axis. The inset shows the temperature derivative of ($\chi \cdot T$) of the data taken at $p$ = 0, 0.2\,GPa and 0.4\,GPa in a narrow temperature window using the same color code.}
	\label{Fig4}
\end{figure}

 A qualitatively different behavior is visible in FIG.\,4 where we show $\chi_{\perp}$ of a crystal with $x$ = 0.105 (\#4) under varying pressure $p \leq$ 0.4\,GPa. The data at $p$ = 0 reveal a sharp kink at $T_{\text{N}}$ = 47.3\,K followed by a rapid decrease down to lower temperatures. Apart from the small feature around 25\,K and the mild upturn below about 10\,K, the data are very similar to that obtained for crystal \#5 (not shown) sharing the same Ge concentration of $x$ = 0.105 within the resolution of the EDX analysis \cite{Peters2022cryst}. On increasing the pressure to $p$ = 0.1\,GPa and beyond, however, marked differences in the characteristics of $\chi_{\perp}$ become apparent. These include the onset of a pronounced upturn in $\chi_{\perp}$ at low temperatures, which is absent for $p$ = 0, and a progressive rounding of the peak in $\chi_{\perp}$ with increasing pressure. The broadening becomes particularly clear for the data taken at 0.3 and 0.4\,GPa, which bear all the characteristics of the valence-change crossover revealed for the crystals with $x$ = 0 and 0.058 (FIG.\,3), i.e., the slightly rounded pronounced drop in the susceptibility which lacks an easy-plane anisotropy, the strong pressure dependence of this feature, and a low-temperature upturn which is practically pressure independent. These observations suggest a pressure-induced change in the ground state for the $x$ = 0.105 crystal from afm order at $p$ = 0 to a valence-change-crossover behavior at $p \geq$ 0.1\,GPa. \\

 This notion is further corroborated by evaluating the quantity d$(\chi \cdot T$)/d$T$ and its variation with pressure shown in the inset of FIG.\,4. For $p$ = 0 a single strongly asymmetric peak is observed. As argued in Appendix B, where we compare d$(\chi \cdot T$)/d$T$ with the contribution to the specific heat related to the 4$f$ electrons, this magnetic response is consistent with a mean-field-type magnetic phase transition. In contrast, more symmetric though increasingly broadened behavior is revealed in d$(\chi \cdot T$)/d$T$ at higher pressures $p \geq$ 0.1\,GPa. As demonstrated for $x$ = 0.058 in the inset of FIG.\,3, a symmetric Lorentzian-like behavior in d$(\chi \cdot T$)/d$T$ characterizes the valence-change crossover. As will be discussed below, we attribute the broadening of these symmetric curves for $x$ = 0.105 to sample inhomogeneities, i.e., small variations in the actual Ge concentration, resulting in a convolution of Lorentzian curves, each of which having a slightly different pressure dependence. To account for this broadening, both the low- and high-temperature flanks were fitted by Lorenzian curves and the pressure-induced shifts of each of which were determined. By using the average value we find $T'_{\text{V}}$ = (45 $\pm$ 1.3)\,K at $p$ = 0.1\,GPa and d$T'_{\text{V}}$/d$p$ = +(90 $\pm$ 10)\,K/GPa for \#4. Practically identical behavior, i.e., a mean-field-type phase transition anomaly in d$(\chi \cdot T$)/d$T$ that develops into broadened Lorentzian-type curves under pressure, was found for the second crystal (\#5) with $x$ = 0.105 in measurements performed at $p$ = 0, 0.1, 0.2 and 0.4\,GPa, see Table 1 for the characteristic temperatures and their pressure dependence. \\

The data in FIG.\,4 suggest that hydrostatic pressure as small as $p$ = 0.1\,GPa is sufficient to induce a drastic change in the ground state for crystals with $x$ = 0.105. To explore this interesting part of the $T-p$ phase diagram in more detail, we show in FIG.\,5 a series of $\chi_{\parallel}$ data plotted as d$(\chi \cdot T$)/d$T$ for temperatures $T \leq$ 80\,K at varying pressure $p \leq$ 0.1\,GPa. By using $\chi_{\parallel}$ as a probe for this investigation, we take advantage of the anisotropy that characterizes the antiferromagnetic order at $p$ = 0. As shown in Ref.\,\cite{Peters2022cryst}, $\chi_{\parallel}$ is almost $T$-independent in the magnetically ordered state, reflecting an easy-plane anisotropy below $T_{\text{N}}$, whereas it shows a rapid drop in the valence-change-crossover regime. As a result, $\chi_{\parallel}$ is more sensitive than $\chi_{\perp}$ for probing a pressure-induced change from magnetic order to valence-change crossover.\\

The data in FIG.\,5 reveal practically identical behavior in d$(\chi_{\parallel} \cdot T$)/d$T$ for $p$ = 0 and 0.03\,GPa, yielding an almost $T$-independent behavior for $T \leq T_{\text{N}}$ = 47.3\,K and a sharp step-like change at $T_{\text{N}}$ (up arrow in FIG.\,5). This behavior reflects a mean-field-type phase transition into afm order (see Appendix B). As there is no identifiable shift in $T_{\text{N}}$ within the experimental resolution on increasing pressure from $p$ = 0 to 0.03\,GPa, we estimate an upper bound for the pressure dependence for the magnetic ordering temperature of d$T_{\text{N}}$/d$p \leq$ +(1 $\pm$ 0.5)\,K/GPa. This small upper limit of the pressure dependence characterizing $T_{\text{N}}$ is also consistent with the data at $p$ = 0.05\,GPa, which essentially show the same behavior as for $p \leq$ 0.03\,GPa at the step-like change, with some small deviations becoming visible on its low- and high-temperature side. By a mild increase of the pressure to 0.06\,GPa, however, the shape of the anomaly in d$(\chi_{\parallel} \cdot T$)/d$T$ changes noticeably in developing a rounded peak and by progressively adopting a more symmetric Lorentzian-like shape. By identifying the position of the maximum with the valence-change-crossover temperature $T'_{\text{V}}$ (down arrow in FIG.\,5), the data in FIG.\,5 yield a pressure dependence of d$T'_{\text{V}}$/d$p$ = +(50 $\pm$ 20)\,K/GPa, which is of the same magnitude as the d$T'_{\text{V}}$/d$p$ values revealed for the $x$ = 0 and 0.058 compounds, cf.\,Table 1. This pressure-induced alteration in the character of the anomaly is accompanied by the appearance of the low-temperature upturn in the susceptibilities $\chi_{\perp}$ (see FIG.4) and $\chi_{\parallel}$ (not shown).

\subsection{$T-p$ phase diagrams}

\begin{figure}[h]
	\includegraphics[width=1.0\columnwidth]{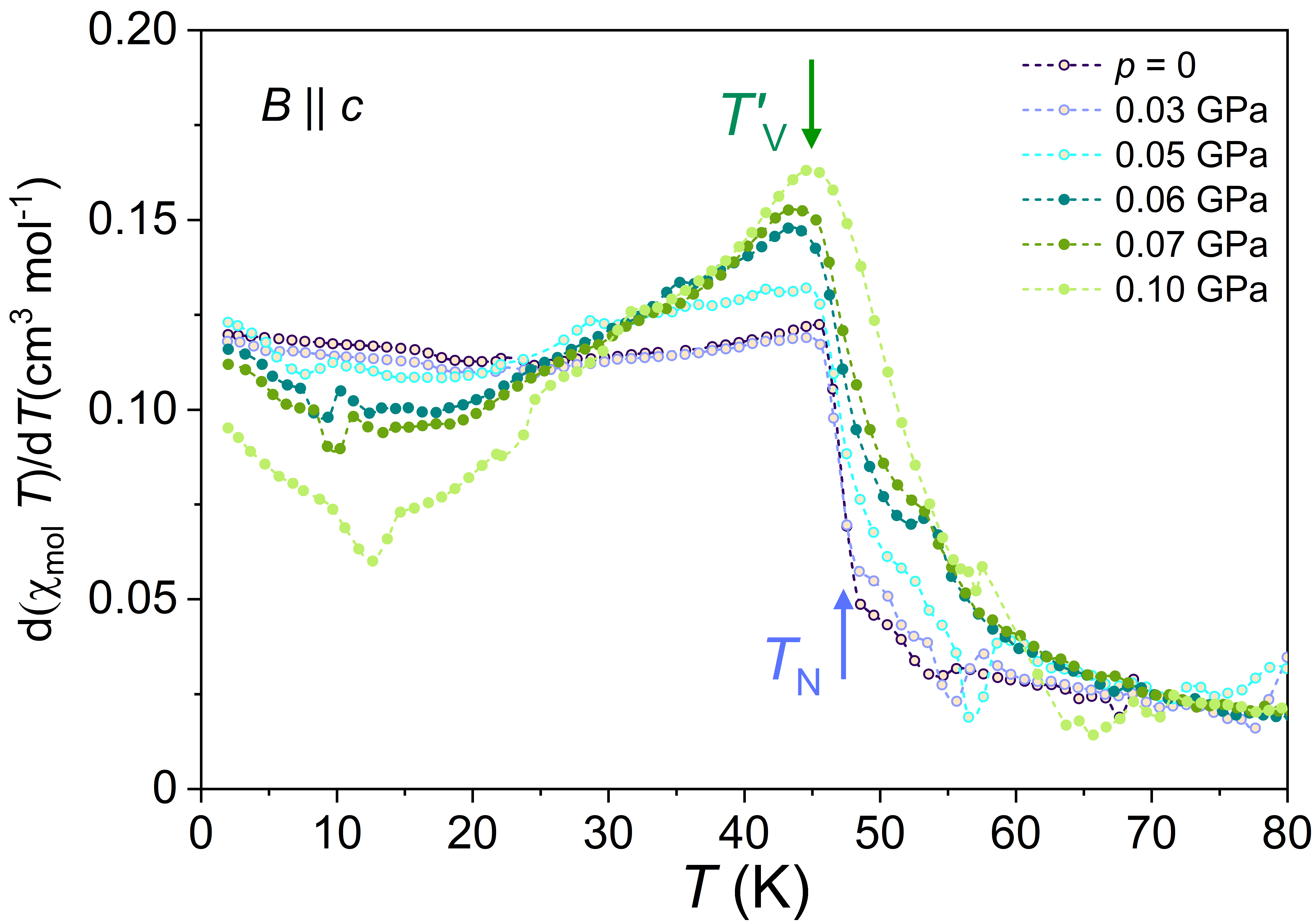}
	\caption{Temperature derivative d($\chi_{\parallel} \cdot T$)/d$T$ of single crystalline EuPd$_2$(Si$_{1-x}$Ge$_x$)$_2$ with $x$ = 0.105 (\#4) for varying small pressures 0 $\leq p \leq$ 0.1\,GPa. The external field of $B$ = 1\,T was oriented along the tetragonal $c$ axis. The position of the magnetic phase transition at $T_{\text{N}}$ is marked by a blue up arrow, whereas the valence-change-crossover temperature $T'_{\text{V}}$ is indicated by a green down arrow.}
	\label{Fig5}
\end{figure}


The anomalies associated with $T'_{\text{V}}$ revealed in FIGs.\,3 and 4 are compiled in a $T-p$ phase diagram (FIG.\,6) of EuPd$_2$(Si$_{1-x}$Ge$_x$)$_2$ for $x$ = 0.058 and $x$ = 0.105. The diagram includes a range of negative pressure (grey area), not accessible by hydrostatic-pressure studies. For $x$ = 0.058 the figure shows the valence-crossover temperature $T'_{\text{V}}$ and its evolution with pressure for $p \leq$ 0.47\,GPa \footnote{Note that the susceptibility data of x = 0.058 (\#3) for $p$ = 0.43\,GPa and $p$ = 0.47\,GPa were omitted in the inset of FIG.\,3 for clarity.}. The figure also shows the width of the anomaly in d($\chi \cdot T$)/d$T$, determined from the full width at half maximum value $\Gamma$ of the Lorentzian curves fitted to the d($\chi \cdot T$)/d$T$ data. In the pressure range investigated the so-derived values for $T'_{\text{V}}$ as well as the lower and upper bounds of $\Gamma$ all show to a good approximation a linear variation with pressure. This enables us to use a linear extrapolation to negative pressures (broken lines) in FIG.\,6. For the $x$ = 0.058 crystal, we find that these lines merge in a single point around $p_{\text{cr}}$ = -(0.63 $\pm$ 0.05)\,GPa and $T_{\text{cr}}$ = (42 $\pm$ 3)\,K. This point may serve as a good approximation for the location of the second-order CEP, shown as the grey filled circle in the main panel of FIG.\,6. By applying a similar procedure to the d($\chi \cdot T$)/d$T$ data for the $x$ = 0.105 crystal (FIG.\,4) for 0.06\,GPa $\leq p \leq$ 0.4\,GPa results in a crossing point at $p_{\text{cr}}$ = -(0.15 $\pm$ 0.05)\,GPa and $T_{\text{cr}}$ = (23 $\pm$ 3)\,K. Note, that the crossover regime for the $x$ = 0.105 crystal is delimited by the low- and high-temperature flanks of a convolution of Lorentz curves fitted to the broadened anomalies in d$(\chi \cdot T$)/d$T$ shown in the inset of FIG.\,4. Despite the more complex behavior for the $x$ = 0.105 compound, resulting from the appearance of magnetic order at low pressures, we may conclude from the trend revealed in FIG.\,6 that by increasing the Ge concentration in EuPd$_2$(Si$_{1-x}$Ge$_x$)$_2$, the CEP of the valence transition is shifted to lower temperatures and is moved closer to $p$ = 0.

\begin{figure}[h]
	\includegraphics[width=1.0\columnwidth]{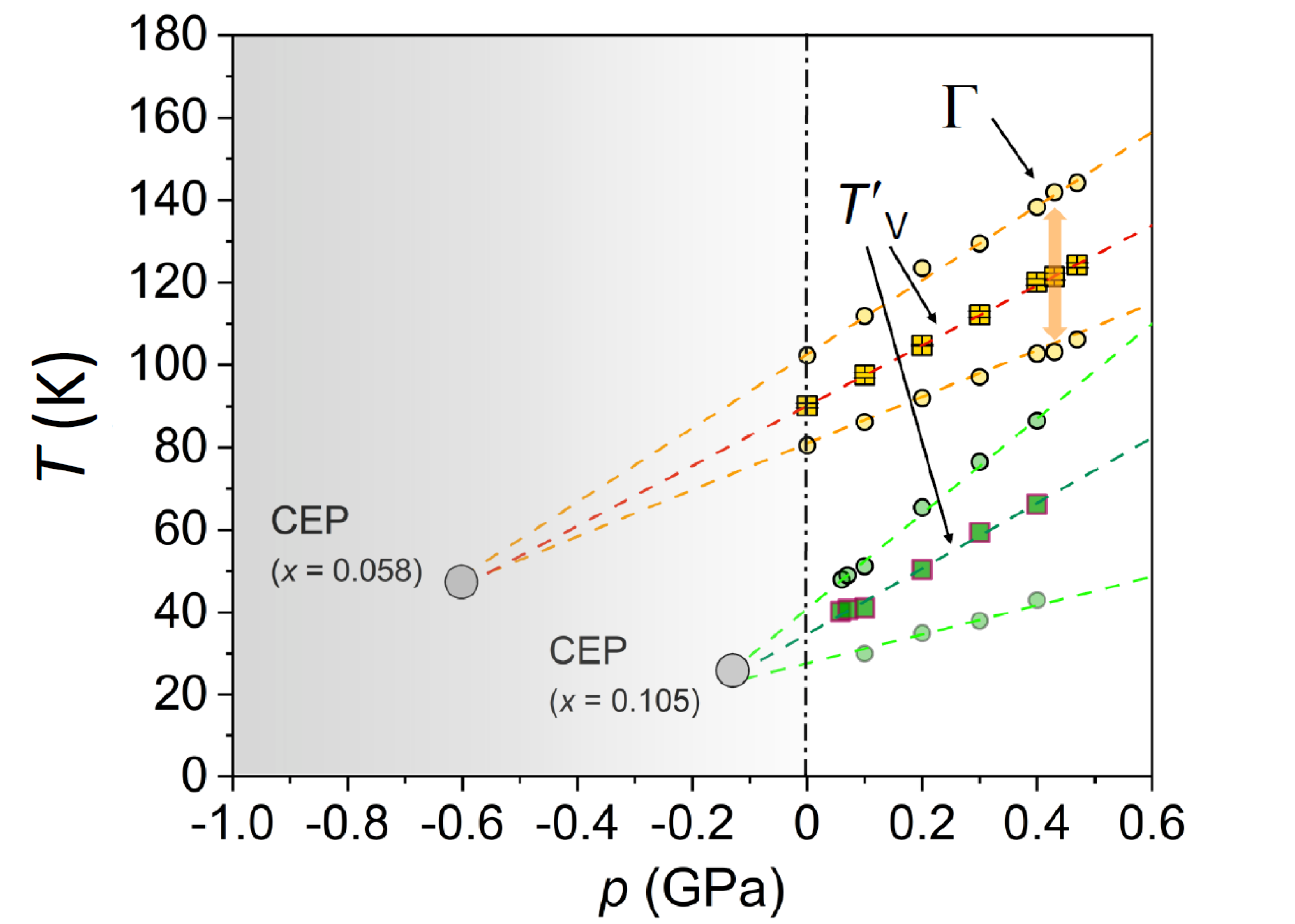}
	\caption{$T-p$ phase diagram of EuPd$_2$(Si$_{1-x}$Ge$_x$)$_2$ with $x$ = 0.058 and $x$ = 0.105 constructed by analyzing d($\chi \cdot T$)/d$T$ data of crystals \#3 and \#4, respectively. The range of negative pressure is indicated in grey. The width of the crossover regime is visualized by pairs of filled circles, the distance of which (indicated by the double arrow) corresponds to the full width at half maximum value $\Gamma$ of the Lorentzian curves fitted to the d($\chi \cdot T$)/d$T$ data. The broken and dotted lines are used to extrapolate anomalies related to $T'_{\text{V}}$ and to the width of the crossover line $\Gamma$ to negative pressures. The point of intersection of these lines (large grey sphere) is considered as a good approximation for the location of the critical endpoint (CEP).}
	\label{Fig1}
\end{figure}

Figure 7 shows details of the $T-p$ phase diagram for the $x$ = 0.105 crystal (\#4) in a narrow pressure window 0 $\leq p \leq$ 0.25\,GPa. The figure includes the region of long-range afm order (blue shaded are) below $T_{\text{N}}$, which is practically pressure independent on the pressure scale shown here. At pressures of $p \geq$ 0.06\,GPa, where no indications for magnetic order can be revealed for $T \geq$ 2\,K, the material is in the valence-crossover region, visualized by the crossover line $T'_{\text{V}}$. In contrast to $T_{\text{N}}$, this crossover line $T'_{\text{V}}$ reveals a strong pressure dependence. The data obtained for crystal \#5 with the same $x$ = 0.105 were found to be fully consistent with the results shown in the phase diagram in FIG.\,7. The figure highlights the exceptional character of the $x$ = 0.105 crystal in showing an extraordinarily high sensitivity of its ground state to hydrostatic pressure.

\begin{figure}[h]
	\includegraphics[width=1.0\columnwidth]{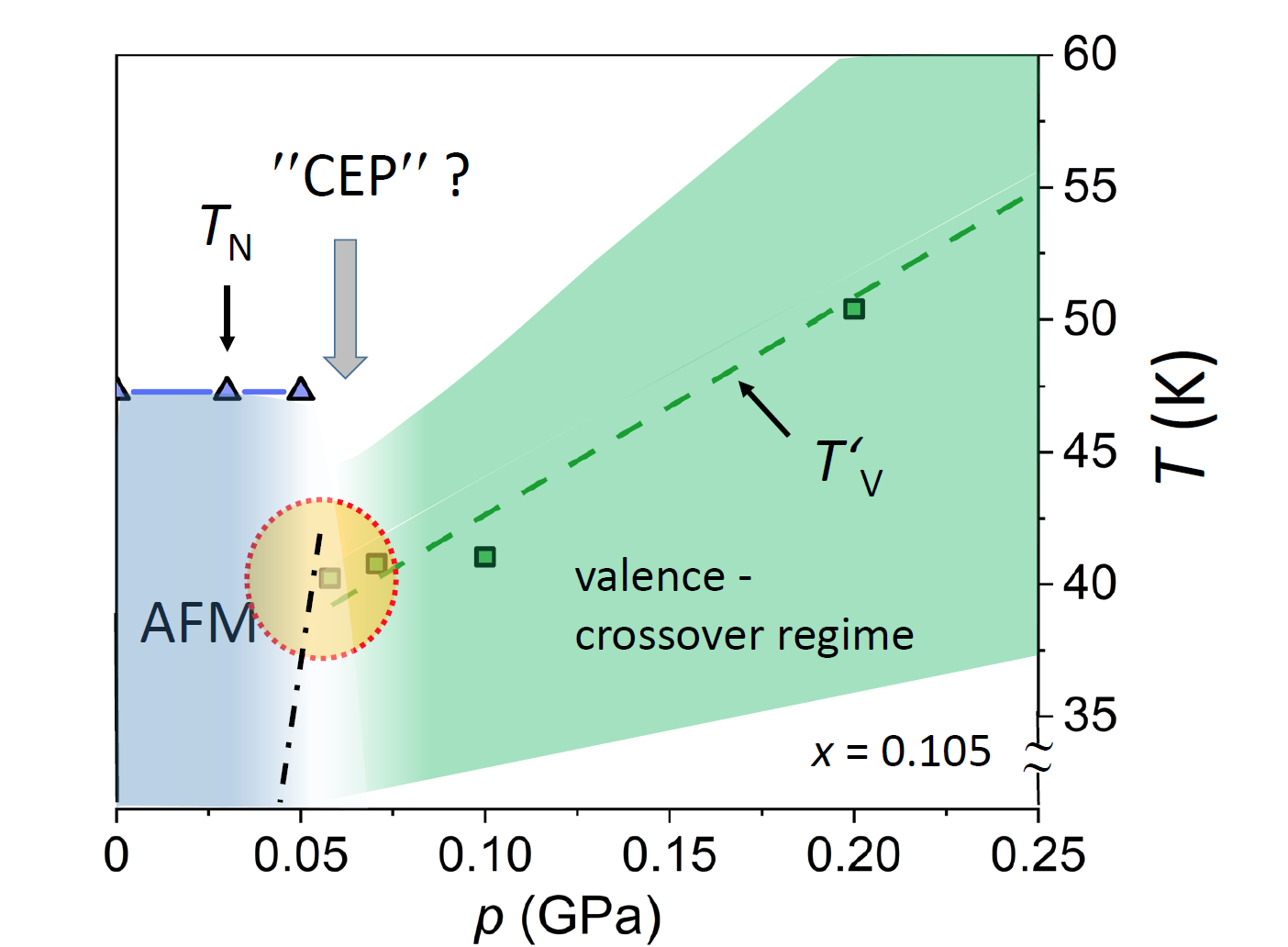}
	\caption{Details of the $T-p$ phase diagram of EuPd$_2$(Si$_{1-x}$Ge$_x$)$_2$ with $x$ = 0.105 based on the analysis of d($\chi \cdot T$)/d$T$ data of crystal \#4. The figure shows the antiferromagnetic (AFM) phase (blue shaded area) below $T_{\text{N}}$ and the valence-crossover regime (green shaded region) visualized by the crossover temperature $T'_{\text{V}}$ and the width of Lorentzian-curves fitted to the d($\chi \cdot T$)/d$T$ data. The black broken line separating the two phases serves as a guide to the eyes. As argued in the main text, this broken line is likely representing a first-order phase transition terminating in a second-order critical endpoint, referred to as "CEP" (red circle). According to the Clausius-Clapeyron equation, see section D of the discussion, we expect a positive pressure dependence of the first-order phase transition line. }
	\label{Fig1}
\end{figure}

\section{Discussion}
\subsection{Effects of physical pressure, Ge substitution and disorder on the valence-change crossover}

The present investigations on single crystalline EuPd$_2$(Si$_{1-x}$Ge$_x$)$_2$ with $x$ = 0, 0.058, and 0.105 highlight the strong sensitivity of the valence-crossover behavior to various parameters. These include the Ge-concentration $x$, the effect of hydrostatic pressure as well as the influence of disorder. The latter effect manifests itself particularly clearly in the marked sample-to-sample variations revealed for crystals \#1 and \#2 with $x$ = 0 (cf. FIG.\,1). Before discussing the effects of pressure and Ge-substitution, we start by addressing the influence of disorder.\\

As shown in Ref.\,\cite{Kliemt2022cryst} for the $x$ = 0 crystals, there are small variations in the Pd:Si ratio along the growth direction of the crystals with the Pd site (Wyckoff position 4$d$) being partially occupied by up to 3\% Si while for all samples the 4$e$ Wyckoff position is completely occupied with Si. As a result of this site-exchange type of disorder in the Pd-Si layers, there are small changes in the regular Si positions, and correspondingly, the bond lengths \cite{Kliemt2022cryst}. The influence of these changes, especially the $a$-axis lattice parameter on $T'_{\text{V}}$, has also been revealed by DFT calculations \cite{Song2023DFT}. Since for the $x$ = 0 crystals all lattice parameters are found to be modified by the Pd:Si ratio \cite{Peters2022cryst}, with the structural changes being at the limit of the experimental resolution in XRD, it is difficult to correlate changes in $T'_{\text{V}}$ to changes of the lattice parameters.\\


This is different for the Ge-substituted crystals where a clear correlation between an increase of the $a$-axis lattice parameter and a decease in $T'_{\text{V}}$, can be observed upon increasing $x$  \cite{Peters2022cryst}. Our observations seem to indicate that sample-to-sample variations are much less pronounced, c.f. the practically identical magnetic behavior revealed for crystals \#4 and \#5 with  $x$ = 0.105. In particular, we find identical behavior for these crystals in the valence-fluctuating regime at $p\geq$ 0.1\,GPa. In these substituted crystals, the Ge atoms because of their size, being significantly larger than Si but similar to Pd, are expected to preferably participate in the site exchange on the Pd position. As a result these Pd-Ge site-exchange processes will only weakly influence the regular Si positions in the surrounding. This favorable side effect of weak Ge substitution may provide a rationale for the lack of significant sample-to-sample variations in $T'_{\text{V}}$ revealed in this study for the crystals with $x >$ 0. \\

As for the effect of pressure, the application of He-gas pressure to single crystals at various substitution levels has demonstrated a strong pressure dependence of $T'_{\text{V}}$ with rates d$T'_{\text{V}}$/d$p$ ranging from +(100 $\pm$ 10)\,K/GPa ($x$ = 0.105) to about +(75 $\pm$ 5)\,K/GPa ($x$ = 0.058). The strong increase of $T'_{\text{V}}$ with pressure is accompanied by a significant widening of the valence-change-crossover range. An extraordinarily high pressure dependence of the valence transition as well as the width of the valence-crossover region are typical characteristics for valence-fluctuating Eu-compounds \cite{Onuki2020phase, Thomson1999pressure}. In the pressure range investigated we find that both $T'_{\text{V}}$ and also the width $\Gamma$ vary linearly with pressure. By using a linear extrapolation to negative pressures these lines merge in a single point. We consider this point as a good approximation for the location of the (hypothetical) second-order CEP out of which crossover regimes are expected to emanate in a V-shaped manner, see, e.g. Ref.\,\cite{Gati2016breakdown}. The data for $x$ = 0.058 and 0.105, which are likely to be less affected by disorder effects as suggested above, indicate that increasing the Ge concentration $x$ corresponds to a shift of the critical pressure by $\Delta p > $ 0. This notion is consistent with the results on polycrystalline samples with $x$ = 0, yielding a critical pressure above 0.7\,GPa \cite{Batlogg1982EuPd2Si2, Schmiester1982EuPd2Si2}. It is tempting to attribute this effect to the negative chemical pressure induced by replacing the smaller Si atoms by the larger (isoelectronic) Ge atoms, corresponding to a widening of the lattice, see also Ref.\,\cite{Peters2022cryst}.

\subsection{Low-temperature increase in susceptibility for $x$ = 0 and 0.058}

The results of the magnetic susceptibility shown in FIGs.\,3 and 4 demonstrate that there is a strong increase in $\chi$($T$) at low temperatures $T \leq$ 25\,K for the crystals showing valence-change-crossover behavior. In contrast to the crossover temperature $T'_{\text{V}}$, which is strongly pressure dependent, this low-temperature upturn is practically unaffected by pressure. Furthermore, no effect is found upon increasing the field from 1\,T to 5\,T (not shown). These observations together with the fact that this upturn is very similar for all crystals investigated makes an interpretation in terms of an impurity contribution very unlikely. Rather it points to an origin which is intrinsic to the state below $T'_{\text{V}}$. This interpretation is consistent with experimental results obtained on various valence-fluctuating Ce- and Yb-compounds discussed in Ref.\,\cite{Lawrence1981valence} and the references cited therein.
It was found that with decreasing $T'_{\text{V}}$ the value of $\chi$($T = 0$) increases. A similar correlation is also observed here in EuPd$_2$(Si$_{1-x}$Ge$_x$)$_2$ for $x < 0.105$. Further systematic studies on the low-temperature upturn are required for substantiating its intrinsic nature.


\subsection{Antiferromagnetic order for $x$ = 0.105}

As discussed in detail in Ref.\,\cite{Peters2022cryst}, the magnetic signatures revealed for $x$ = 0.105 at $p$ = 0, including a sharp kink at 47.3\,K followed upon cooling by an easy-plane anisotropy, and a mean-field-type phase transition in the specific heat, provide clear evidence for long-range afm order. This easy-plane anisotropy is also visible in the quantity d($\chi \cdot T$)/d$T$ as displayed in {FIGs. 5 and 9.} The observed $T_{\text{N}}$ and its small pressure dependence revealed in the present work fall into the ranges typically observed for intermetallic compounds featuring divalent Eu ions, see Refs.\,\cite{Anand2015AFM, Onuki2020phase, Hossain2004VT, Baenitz2006NMR, Dionicio2006VT, Hossain2004AFM, Muthu2016phase, Nakamura2015trans, Gouchi2020AFM, Fukuda2003} and references cited therein. In this context we also mention the magnetic anisotropy revealed in the paramagnetic state for $x$ = 0.105 (see inset of FIG.\,2), which is often seen in Eu$^{2+}$-based magnets, especially in the 122 compounds \cite{Anand2015AFM}.

\subsection{Pressure-induced change from antiferromagnetic order to valence-crossover behavior for $x$ = 0.105}

The $T-p$ phase diagram in FIG.\,7 for $x$ = 0.105 indicates a drastic change in the material's ground state from afm order to valence crossover in response to a tiny increase in the applied pressure from 0.03\,GPa to 0.06\,GPa. The interplay between afm order and valence fluctuations was discussed in a theoretical work by S.\,Watanabe and K.\,Miyake \cite{Watanabe2011Theo}. It was found that for strong valence fluctuations, the valence-change-crossover regime is separated from the magnetic state by a first-order phase transition. To explore the possibility of such a $p$-induced first-order transition in EuPd$_2$(Si$_{1-x}$Ge$_x$)$_2$ for $x$ = 0.105 in more detail, we attempted to gain access to the entropy contribution of the 4$f$ electrons, $S_{\text{4}f}$, for the $x$ = 0.105 compound and to explore the variation of $S_{\text{4}f}$ with pressure. To this end we again take advantage of the proportionality between d($\chi \cdot T$)/d$T$ and the 4$f$-related electronic specific heat (see FIG.\,10 in Appendix B for details). By using the experimental data shown in FIGs.\,3, 4, 5, and 9 and by integrating the expression [(1/$T$) $\cdot$ d($\chi \cdot T$)/d$T$] $\propto$ $C_{\text{4}f}$, we get an estimate of the entropy related to the 4$f$ electrons. An estimate of the $C_{\text{4}f}$($T$) contribution was obtained in Ref.\,\cite{Peters2022cryst}, (see also Appendix B for details.)


\begin{figure}[h]
	\includegraphics[width=1.0\columnwidth]{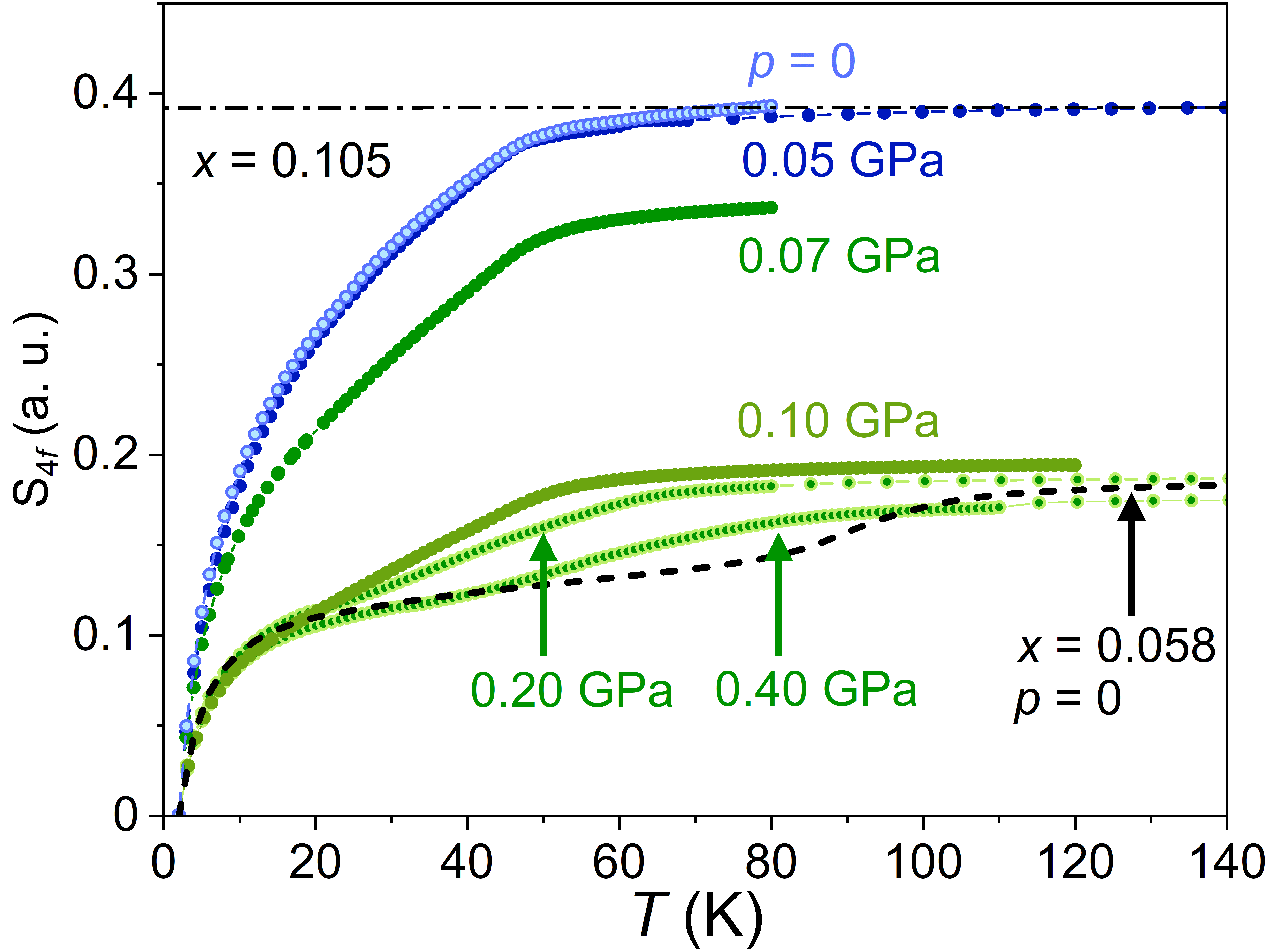}
	\caption{Estimate of the entropy, $S_{\text{4}f}$, related to the 4$f$ electrons of \#4 with $x$ = 0.105 in arbitrary units obtained by integrating [(1/$T$) $\cdot$ d($\chi \cdot T$)/d$T$] $\propto$ $C_{\text{4}f}$. See Appendix B for details. The light blue (full blue) circles represent $S_{\text{4}f}$ at $p$ = 0 (0.05\,GPa), where the system adopts an antiferromagnetically-ordered ground state. The dashed dotted line marks the maximum entropy $S_{\text{max}}$, which is expected to be a sizable fraction of $R \cdot$ ln(2$S$ + 1) of the $S$ = 7/2 state. The dark green circles correspond to $S_{\text{4}f}$ at $p$ = 0.07\,GPa. The data obtained for $p$ = 0.1, 0.2, and 0.4\,GPa (full light green circles) show a significantly reduced $S_{\text{4}f}$ approaching about 0.5 $\cdot S_{\text{max}}$. The black dashed line corresponds to $S_{\text{4}f}$ of the $x$ = 0.058 crystal (\#3) at $p$ =0 which shows valence-crossover behavior.}
	\label{app_II}
\end{figure}

The so-derived $S_{\text{4}f}$($T$) is shown in FIG.\,8 for varying pressure $p\leq$ 0.4\,GPa. The figure indicates that at $p$ = 0, where the system orders antiferromagnetically below $T_{\text{N}}$ = 47.3\,K, $S_{\text{4}f}$ tends to level off at its highest value $S_{\text{max}}$. Practically identical behavior within the experimental resolution is obtained for $p$ = 0.05\,GPa. On mildly increasing the pressure to 0.07\,GPa, however, a significant reduction in the maximum entropy to $0.87 \cdot S_{\text{max}}$ is observed. For pressures of 0.1, 0.2 and 0.4\,GPa, where the system undergoes the valence crossover, the maximum entropy is further reduced to about $0.5 \cdot S_{\text{max}}$. The data in FIG.\,8 suggest that crossing the boundary between afm order and valence crossover on increasing the pressure at $T$ = const. from 0.05 over 0.07 to 0.1\,GPa is accompanied by a discontinuous change in entropy, indicative of a first-order phase transition. Moreover, according to the Clausius-Clapeyron equation d$T^{*}$/d$p$ = $\Delta V$/$\Delta S$, a negative volume change $\Delta V <$ 0 together with the drop in entropy $\Delta S < $ 0 on going from the antiferromagnetically-ordered state to the valence-change-crossover regime, correspond to a positive pressure dependence of the corresponding first-order phase transition line $T^{*}$. In order to verify the first-order character of the transition, continuous pressure sweeps at $T$ = const. conditions of a thermodynamic probe would be desirable. Whereas pressure sweeps in combination with measurements of the specific heat or magnetization are very challenging, they are feasible for measurements probing the relative length changes \cite{Manna2012druck, Agarmani2022druck}. As was demonstrated in Refs.\,\cite{Gati2016breakdown, Wolf2022SCES} these experiments are very sensitive to the character of the phase transition and the presence of valence fluctuations and are thus considered key experiments for exploring this part of the phase diagram in detail.\\

An exciting implication of such a first-order phase transition would be the presence of a particular type of critical endpoint, referred to as "CEP", where the  valence-change crossover emerges directly out of an antiferromagnetically-ordered state. In contrast to the CEP for canonical valence-fluctuating systems, where effects of interacting charge- and lattice degrees of freedom are expected, here an additional degree of freedom resulting from the nearby magnetic order comes into play. As a result we may expect strong-coupling effects between spin-, charge- and lattice degrees of freedom upon approaching this "CEP". According to the phase diagram depicted in FIG.\,7, we locate the critical endpoint for $x$ = 0.105 at $p_{\text{cr}} \approx$ 0.06\,GPa and $T_{\text{cr}}$ $\approx$ 39\,K. \\

\section{Conclusions} 
The interplay between valence-change crossover ($x \leq$ 0.058) and magnetic order ($x$ = 0.105) has been investigated by measurements of the magnetic susceptibility on single crystalline EuPd$_2$(Si$_{1-x}$Ge$_x$)$_2$ under varying hydrostatic (He-gas) pressure $p \leq$ 0.5\,GPa. At ambient pressure, the crystals with $x$ = 0 and 0.058 show valence-change-crossover behavior with a crossover temperature $T'_{\text{V}}$ ranging from 90\,K to about 155\,K. $T'_{\text{V}}$ is characterized by a strong pressure dependence of typically d$T'_{\text{V}}$/d$p$ = +(80 $\pm$ 10)\,K/GPa. On the other hand, for $x$ = 0.105, as a consequence of negative chemical pressure, long-range afm order is observed below $T_{\text{N}}$ = 47.3\,K. In contrast to the strong pressure dependence of $T'_{\text{V}}$, $T_{\text{N}}$ remains practically unaffected by pressure for $p \leq$ 0.05\,GPa. On further increasing the pressure to 0.07 and 0.1\,GPa, however, the $x$ = 0.105 crystal changes its ground state from afm order to valence-change crossover. Estimates of the entropy contribution related to the 4$f$ electrons indicate this transition to be of first order. Our results suggest the existence of a special second-order critical endpoint for $x$ = 0.105 at $p_{\text{cr}} \approx$ 0.06\,GPa and $T_{\text{cr}}$ $\approx$ 39\,K. Unlike the critical endpoint that terminates a first-order valence-transition line $T_{\text{V}}$, this endpoint distinguishes itself by valence-change-crossover behavior emerging directly out of an antiferromagnetically-ordered state. As a result, strong-coupling effects between fluctuating charge-, spin-, and lattice degrees of freedom can be expected. The low value of $p_{\text{cr}}$, conveniently accessible by He-gas-pressure experiments, makes this system a well-suited target material for detailed investigations of such strong-coupling effects.

\section*{Acknowledgments}
We thank Ch. Geibel for fruitful discussions. This work was supported by the Deutsche Forschungsgemeinschaft (DFG, German Research Foundation) through TRR 288 - 422213477 (projects A01 and A03).

\appendix
\begin{appendix}

\section{Magnetic anisotropy}

In order to estimate the electronic entropy $S_{\text{4}f}$ from d($\chi \cdot T$)/d$T$ data (see subsection B of the Appendix below), the anisotropy in the magnetic response has to be taken into account. Figure 9 shows the temperature derivative d($\chi_{\perp} \cdot T$)/d$T$ for $x$ = 0.105 (\#4) for varying small pressures 0 $\leq p \leq$ 0.1\,GPa. In contrast to the corresponding plot for $\chi_{\parallel}$ (FIG.\,5), the distinction between anomalies associated to $T_{\text{N}}$ and $T'_{\text{V}}$ are less clear for $\chi_{\perp}$. The reason for that lies in the magnetic easy-plane anisotropy that develops below $T_{\text{N}}$ see Ref. \cite{Peters2022cryst} for $\chi_{\perp}$ and $\chi_{\parallel}$ data, manifesting itself in a rapidly decreasing $\chi_{\perp}$ as opposed to a practically $T$-independent $\chi_{\parallel}$. Since a decreasing $\chi_{\perp}$ is also revealed upon cooling through $T'_{\text{V}}$ makes it more difficult to discriminate between both scenarios. Despite these difficulties, a clear statement can be made by analyzing the data on the high-temperature flank of the anomaly. For $0 \leq p \leq 0.05$\,GPa the d($\chi_{\perp} \cdot T$)/d$T$ data exhibit a step-like change, reminiscent of the mean-field-like phase transition anomaly observed in the specific heat at $T_N$ (FIG.\,10) without any resolvable pressure dependence. On further increasing the pressure to 0.07 and 0.1\,GPa, however, a shoulder develops in d($\chi_{\perp} \cdot T$)/d$T$ above 45\,K, indicating a significant pressure dependence of this contribution. The data are consistent with a pressure-induced change from afm order for $p \leq$ 0.05\,GPa to valence-change-crossover behavior for $p \geq$ 0.07\,GPa -- the same conclusion as drawn by analyzing corresponding $\chi_{\parallel}$ data in FIG.\,5.

\begin{figure}[h]
	\includegraphics[width=1.0\columnwidth]{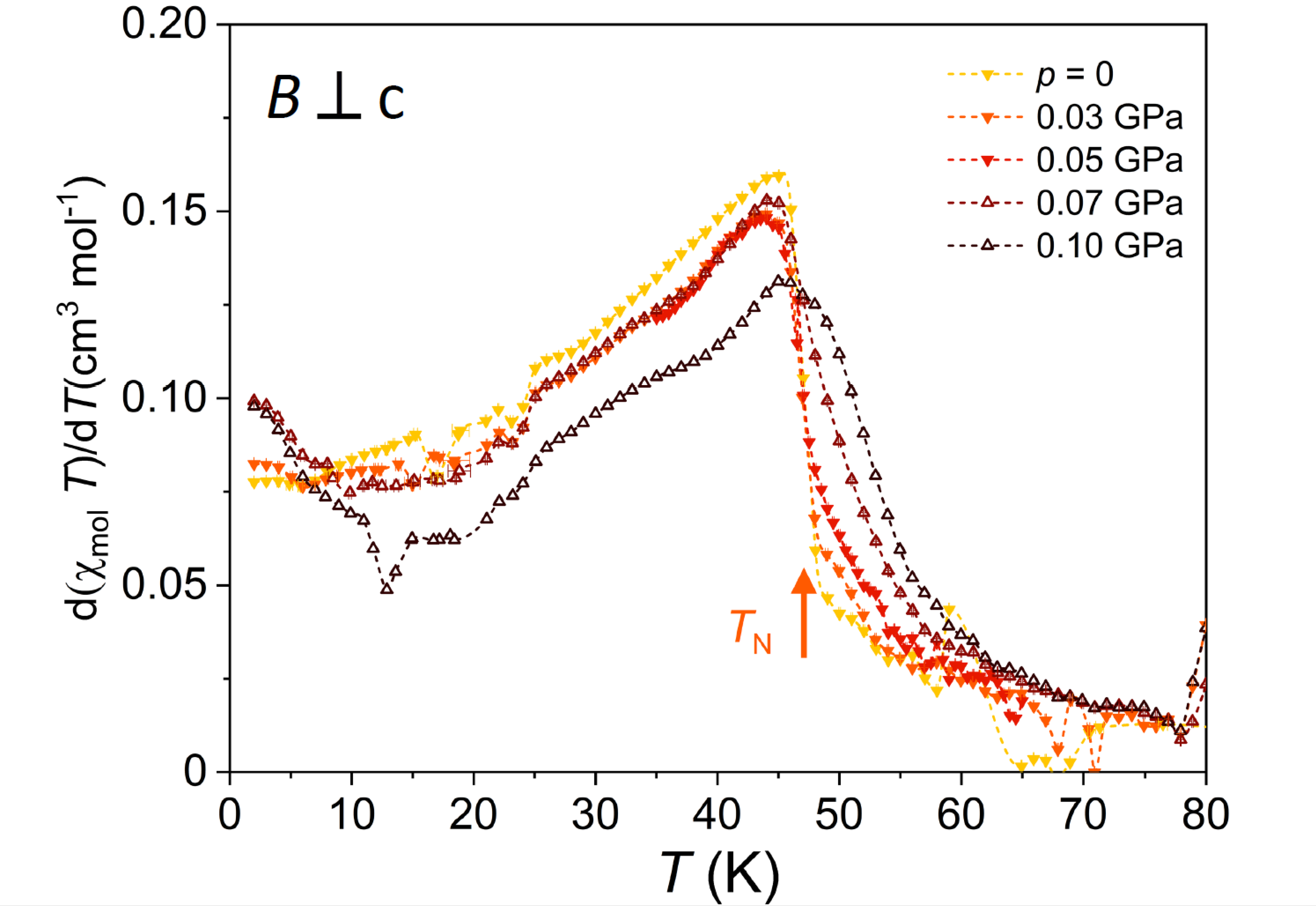}
	\caption{Temperature derivative d($\chi_{\perp} \cdot T$)/d$T$ of single crystalline EuPd$_2$(Si$_{1-x}$Ge$_x$)$_2$ for $x$ = 0.105 (\#4) at varying small pressures 0 $\leq p \leq$ 0.1\,GPa. The external field of $B$ = 1\,T was oriented perpendicular to the $c$ axis. The position of the magnetic phase transition at $T_N$ is marked by a red up arrow.}
	\label{app_I}
\end{figure}

\section{Estimate of 4$f$ entropy}

By using thermodynamic arguments M.E. Fisher showed that for antiferromagnets the quantity d($\chi \cdot T$)/d$T$ is proportional to the magnetic specific heat \cite{Fisher1962specHeat}. Below we demonstrate that for the present EuPd$_2$(Si$_{1-x}$Ge$_x$)$_2$ system the proportionality between electronic contributions to the specific heat and d($\chi \cdot T$)/d$T$ holds true not only for magnetic contributions around $T_{\text{N}}$ but also for the contributions resulting from valence fluctuations around $T'_{\text{V}}$.

To this end we plot the 4$f$ contribution to the specific heat, $C_{\text{4}f}$, for crystals with $x$ = 0.105 (FIG. 10a) and 0.058 (FIG. 10b) (right scales) along with the corresponding d($\chi \cdot T$)/d$T$ data (left scales). $C_{\text{4}f}$ was obtained from the measured specific heat by subtracting the lattice contributions and a small non-4$f$-related electronic contribution as described in detail in Ref.\cite{Peters2022cryst}. The lattice contribution was obtained by a Debye fit using two Debye temperatures \cite{Peters2022cryst}. To account for the non-4$f$-related electronic contribution, i.e., the contributions of Eu 5$d$, Pd 4$d$ and Si 3$p$ states at the Fermi level, a term $\gamma \cdot T$ was used with $\gamma$ = 6 mJ/molK$^2$, found for the reference material LaPd$_2$Si$_2$ which lacks 4$f$ electrons \cite{Peters2022cryst}. 


Figures 10a and 10b highlight the clear correspondence in the shape of the observed features, yielding asymmetric, mean-field-type phase transition anomalies in both quantities at the magnetic transition for $x$ = 0.105 (FIG.\,10a) as opposed to symmetric Lorentzian-shaped curves characterizing the valence-change crossover at $T'_{\text{V}}$ (FIG.\,10b), see also Ref. \cite{Hossain2004AFM} for corresponding C(T) results on EuNi$_2$(Si$_{1-x}$Ge$_x$)$_2$. After having established this correspondence between d($\chi \cdot T$)/d$T$ and the 4$f$-related electronic specific heat for these well-defined borderline cases, we feel confident to use the quantity d($\chi \cdot T$)/d$T$ also for identifying pressure-induced changes in the character of the anomalies for the crystal with $x$ = 0.105 (cf.\,FIG.\,4). \\

\begin{figure}[h]
	\includegraphics[width=1.0\columnwidth]{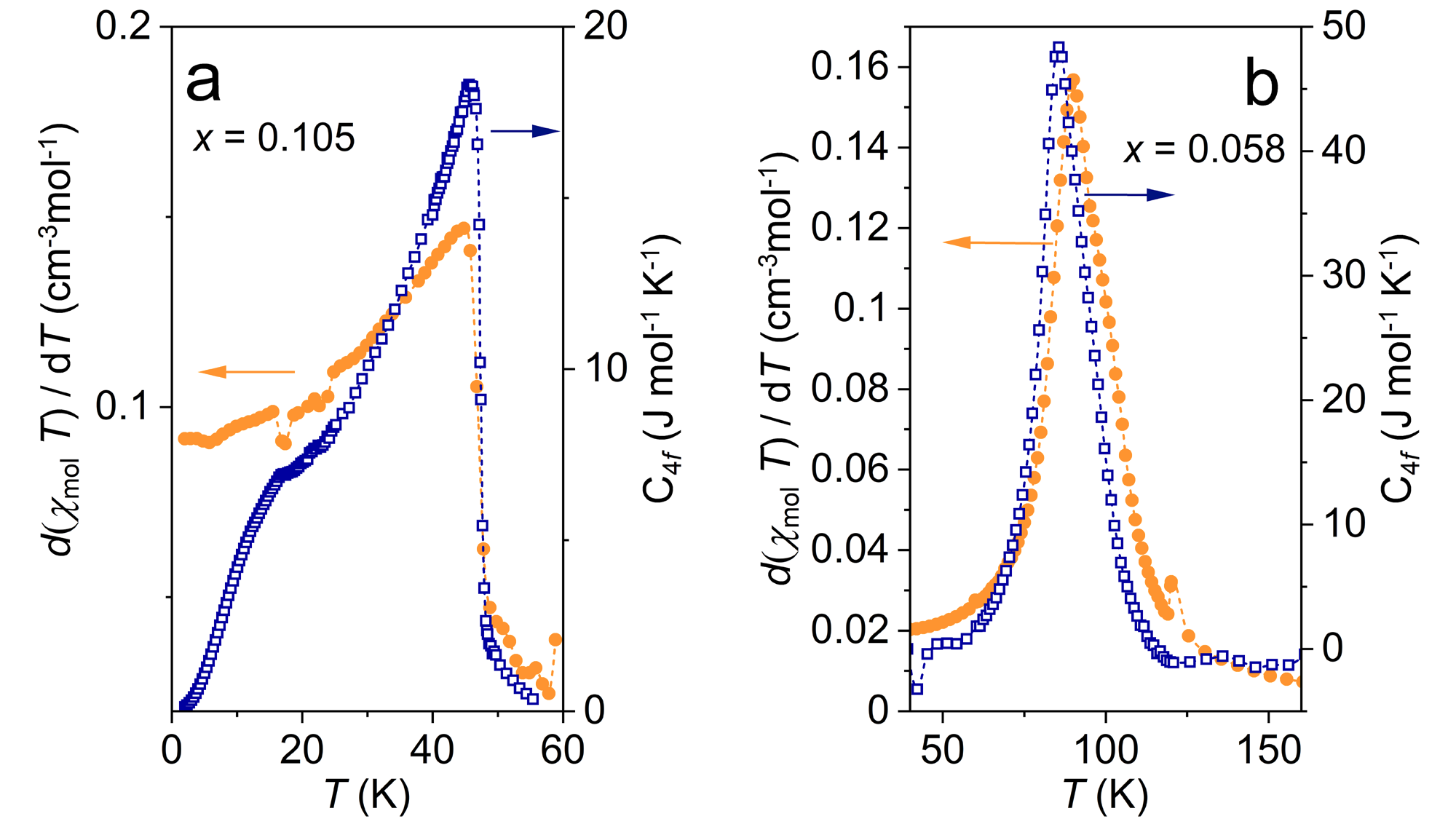}
	\caption{Comparison of the 4$f$-related specific heat (right scales, open blue squares) of EuPd$_2$(Si$_{1-x}$Ge$_x$)$_2$ \cite{Peters2022cryst}, with d($\chi \cdot T$)/d$T$ (left scales, closed orange circles), plotted on the same temperature axes, for $x$ = 0.105 (a) and 0.058 (b). }
	\label{app_II}
\end{figure}

Besides the above qualitative statements, the correspondence between $C_{\text{4}f}$ and d($\chi \cdot T$)/d$T$ can be used also for estimating the 4$f$ entropy $S_{\text{4}f}$. By integrating the expression [(1/$T$) $\cdot$ d($\chi \cdot T$)/d$T$ $\propto$ $C_{\text{4}f}/T$] an estimate of the \textcolor{red}{4$f$} entropy $S_{\text{4}f}$ can be obtained. Note that in order to account for the magnetic anisotropy of the antiferromagnetically-ordered state, we use an expression [d($\chi_{\perp} \cdot T$)/d$T$ + 2 $\cdot$ d($\chi_{\parallel} \cdot T$)/d$T$]/3 to estimate the magnetic entropy as a function of temperature at different pressures. By doing so we are able to follow the evolution of $S_{\text{4}f}$ in small pressure steps. As displayed in FIG.\,8, the so-derived $S_{\text{4}f}$ = $S_{\text{mag}}$ for the crystal with $x$ = 0.105, which orders antiferromagnetically below $T_{\text{N}}$ = 47.3\,K, shows the tendency to saturate at intermediate temperatures at a value $S_{\text{max}}$. A practically identical behavior is obtained on applying pressure of 0.05\,GPa. On further increasing the pressure to 0.07\,GPa, we observe a reduction to about 0.87 $\cdot S_{\text{max}}$. Finally, for pressures of 0.1, 0.2 and 0.4\,GPa, the highest pressure of our experiments, a plateau is reached at around 0.5 of $S_{\text{max}}$. About the same value is reached for the crystal with $x$ = 0.058 at $p$ = 0 (broken line) which shows valence-change-crossover behavior. The drastic change in $S_{\text{4}f}$ on mildly increasing the pressure from 0.05 to 0.07 and finally 0.1\,GPa is considered as a strong indication for a pressure-induced $1^{st}$ order transition on going from the antiferromagnetically ordered state to the valence-change crossover region.

\section{Modeling of d($\chi \cdot T$)/d$T$}

\begin{figure}[h]
	\includegraphics[width=1.0\columnwidth]{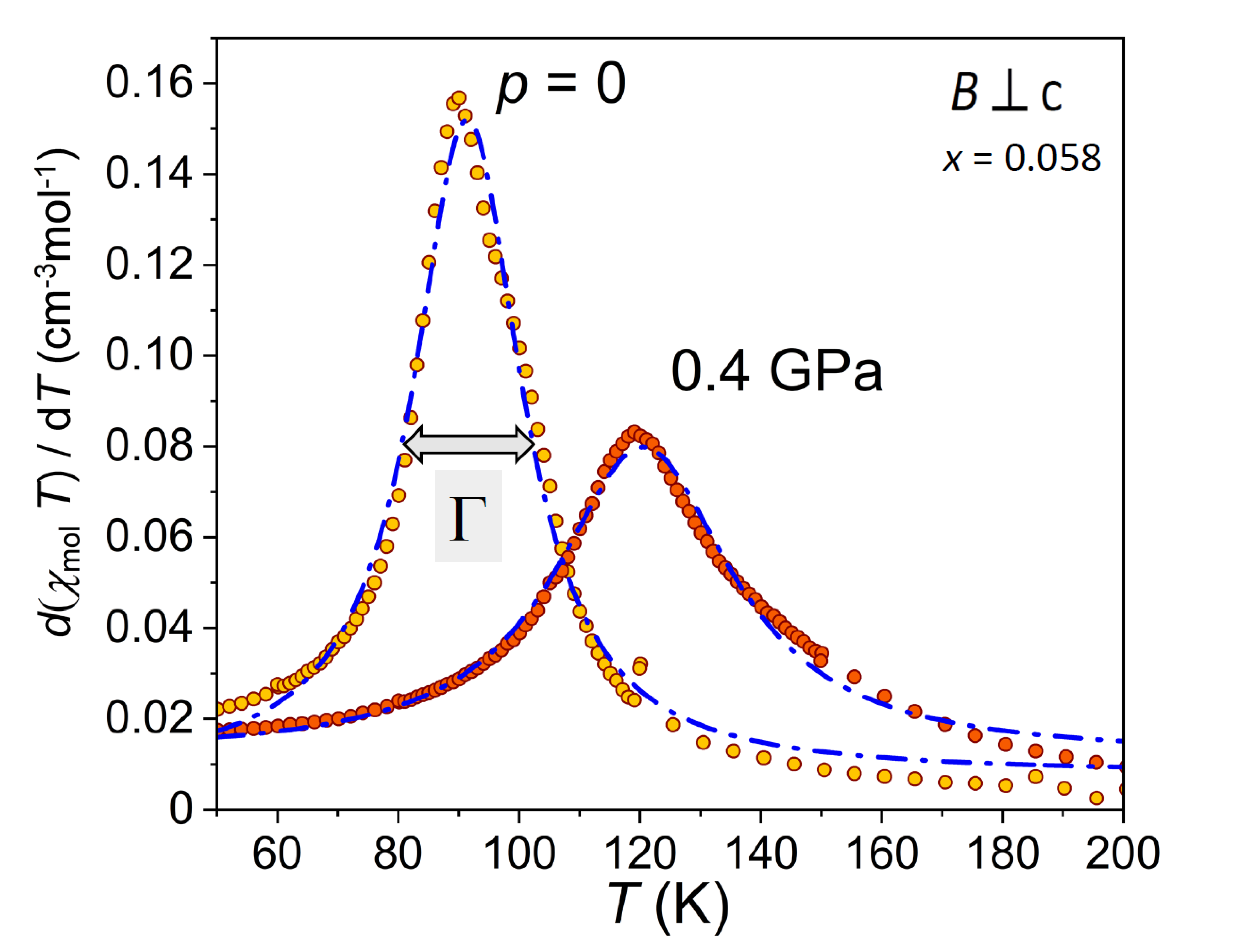}
	\caption{Magnetic susceptibility data plotted as d($\chi \cdot T$)/d$T$ of single crystalline EuPd$_2$(Si$_{1-x}$Ge$_x$)$_2$ with $x$ = 0.058 for crystal \#3. The figure shows the data for $p$ = 0 (full dark yellow circles) together with the data taken at $p$ = 0.4\,GPa (full red circles). The blue broken line is a fit to the data by using a Lorentzian function. The full width at half maximum $\Gamma$ of the Lorentzian is indicated by the grey arrow.}
	\label{Fig1}
\end{figure}
Figure 11 exhibits the temperature dependence of the quantity d($\chi \cdot T$)/d$T$ at $p$ = 0 and $p$ = 0.4\,GPa of EuPd$_2$(Si$_{1-x}$Ge$_x$)$_2$ with $x$ = 0.058 for crystal \#3. As discussed in the main text above, these representations were used for parameterizing the valence-crossover regime and to construct the $p-T$ phase diagram. The figure demonstrates that the Lorentzian curves (broken lines) provide a good fit to the data, capturing the characteristic features, namely the position of the maximum, the symmetric shape of the curves (around the maximum) as well as their width. For the latter, the full width at half maximum $\Gamma$ (double arrow in FIG.\,11) of the Lorentzian is used. The characteristic energy scale of the valence transition or valence crossover, characterized by the position of the maximum of the Lorentzian, corresponds to the energy difference between the Eu$^{2+}$(4$f^7$) and Eu$^{3+}$(4$f^6$) configurations. The corresponding interchange process between these two electronic configurations involves hopping of the localized 4$f$ electrons to conduction band states and vice versa \cite{Wohlleben1975ICF}. This process bears some resemblance of a damped oscillator, described by a Lorentzian function. \\


\end{appendix}

\bibliography{bibly}

\begin{thebibliography}{41}%
\makeatletter
\providecommand \@ifxundefined [1]{%
 \@ifx{#1\undefined}
}%
\providecommand \@ifnum [1]{%
 \ifnum #1\expandafter \@firstoftwo
 \else \expandafter \@secondoftwo
 \fi
}%
\providecommand \@ifx [1]{%
 \ifx #1\expandafter \@firstoftwo
 \else \expandafter \@secondoftwo
 \fi
}%
\providecommand \natexlab [1]{#1}%
\providecommand \enquote  [1]{``#1''}%
\providecommand \bibnamefont  [1]{#1}%
\providecommand \bibfnamefont [1]{#1}%
\providecommand \citenamefont [1]{#1}%
\providecommand \href@noop [0]{\@secondoftwo}%
\providecommand \href [0]{\begingroup \@sanitize@url \@href}%
\providecommand \@href[1]{\@@startlink{#1}\@@href}%
\providecommand \@@href[1]{\endgroup#1\@@endlink}%
\providecommand \@sanitize@url [0]{\catcode `\\12\catcode `\$12\catcode
  `\&12\catcode `\#12\catcode `\^12\catcode `\_12\catcode `\%12\relax}%
\providecommand \@@startlink[1]{}%
\providecommand \@@endlink[0]{}%
\providecommand \url  [0]{\begingroup\@sanitize@url \@url }%
\providecommand \@url [1]{\endgroup\@href {#1}{\urlprefix }}%
\providecommand \urlprefix  [0]{URL }%
\providecommand \Eprint [0]{\href }%
\providecommand \doibase [0]{https://doi.org/}%
\providecommand \selectlanguage [0]{\@gobble}%
\providecommand \bibinfo  [0]{\@secondoftwo}%
\providecommand \bibfield  [0]{\@secondoftwo}%
\providecommand \translation [1]{[#1]}%
\providecommand \BibitemOpen [0]{}%
\providecommand \bibitemStop [0]{}%
\providecommand \bibitemNoStop [0]{.\EOS\space}%
\providecommand \EOS [0]{\spacefactor3000\relax}%
\providecommand \BibitemShut  [1]{\csname bibitem#1\endcsname}%
\let\auto@bib@innerbib\@empty
\bibitem [{\citenamefont {Zacharias}\ \emph {et~al.}(2015)\citenamefont
  {Zacharias}, \citenamefont {Paul},\ and\ \citenamefont
  {Garst}}]{Zacharias2015QCElasticity}%
  \BibitemOpen
  \bibfield  {author} {\bibinfo {author} {\bibfnamefont {M.}~\bibnamefont
  {Zacharias}}, \bibinfo {author} {\bibfnamefont {I.}~\bibnamefont {Paul}},\
  and\ \bibinfo {author} {\bibfnamefont {M.}~\bibnamefont {Garst}},\ }\bibfield
   {title} {\bibinfo {title} {{Q}uantum {C}ritical {E}lasticity},\ }\href
  {http://doi.org/10.1103/PhysRevLett.115.025703} {\bibfield  {journal}
  {\bibinfo  {journal} {Phys. Rev. Lett.}\ }\textbf {\bibinfo {volume} {115}},\
  \bibinfo {pages} {025703} (\bibinfo {year} {2015})}\BibitemShut {NoStop}%
\bibitem [{\citenamefont {Gati}\ \emph {et~al.}(2016)\citenamefont {Gati},
  \citenamefont {Garst}, \citenamefont {Manna}, \citenamefont {Tutsch},
  \citenamefont {Wolf}, \citenamefont {Bartosch}, \citenamefont {Schubert},
  \citenamefont {Sasaki}, \citenamefont {Schlueter},\ and\ \citenamefont
  {Lang}}]{Gati2016breakdown}%
  \BibitemOpen
  \bibfield  {author} {\bibinfo {author} {\bibfnamefont {E.}~\bibnamefont
  {Gati}}, \bibinfo {author} {\bibfnamefont {M.}~\bibnamefont {Garst}},
  \bibinfo {author} {\bibfnamefont {R.~S.}\ \bibnamefont {Manna}}, \bibinfo
  {author} {\bibfnamefont {U.}~\bibnamefont {Tutsch}}, \bibinfo {author}
  {\bibfnamefont {B.}~\bibnamefont {Wolf}}, \bibinfo {author} {\bibfnamefont
  {L.}~\bibnamefont {Bartosch}}, \bibinfo {author} {\bibfnamefont
  {H.}~\bibnamefont {Schubert}}, \bibinfo {author} {\bibfnamefont
  {T.}~\bibnamefont {Sasaki}}, \bibinfo {author} {\bibfnamefont {J.~A.}\
  \bibnamefont {Schlueter}},\ and\ \bibinfo {author} {\bibfnamefont
  {M.}~\bibnamefont {Lang}},\ }\bibfield  {title} {\bibinfo {title} {Breakdown
  of hooke's law of elasticity at the {M}ott critical endpoint in an organic
  conductor},\ }\href@noop {} {\bibfield  {journal} {\bibinfo  {journal}
  {Science Advances}\ }\textbf {\bibinfo {volume} {2}},\ \bibinfo {pages}
  {e1601646} (\bibinfo {year} {2016})}\BibitemShut {NoStop}%
\bibitem [{\citenamefont {Lawrence}\ \emph {et~al.}(1981)\citenamefont
  {Lawrence}, \citenamefont {Riseborough},\ and\ \citenamefont
  {Parks}}]{Lawrence1981valence}%
  \BibitemOpen
  \bibfield  {author} {\bibinfo {author} {\bibfnamefont {J.~M.}\ \bibnamefont
  {Lawrence}}, \bibinfo {author} {\bibfnamefont {P.~S.}\ \bibnamefont
  {Riseborough}},\ and\ \bibinfo {author} {\bibfnamefont {R.~D.}\ \bibnamefont
  {Parks}},\ }\bibfield  {title} {\bibinfo {title} {Valence fluctuation
  phenomena},\ }\href@noop {} {\bibfield  {journal} {\bibinfo  {journal} {Rep.
  Prog. Phys.}\ }\textbf {\bibinfo {volume} {44}},\ \bibinfo {pages} {1}
  (\bibinfo {year} {1981})}\BibitemShut {NoStop}%
\bibitem [{\citenamefont {Ryan}\ \emph {et~al.}(2023)\citenamefont {Ryan},
  \citenamefont {Bud’ko}, \citenamefont {Kuthanazhi},\ and\ \citenamefont
  {Canfield}}]{Ryan2023VTMag}%
  \BibitemOpen
  \bibfield  {author} {\bibinfo {author} {\bibfnamefont {D.~H.}\ \bibnamefont
  {Ryan}}, \bibinfo {author} {\bibfnamefont {S.~L.}\ \bibnamefont {Bud’ko}},
  \bibinfo {author} {\bibfnamefont {B.}~\bibnamefont {Kuthanazhi}},\ and\
  \bibinfo {author} {\bibfnamefont {P.~C.}\ \bibnamefont {Canfield}},\
  }\bibfield  {title} {\bibinfo {title} {Valence and magnetism in
  {E}u{P}d$_3${S}$_4$ and ({Y}, {L}a)$_x${E}u$_{1-x}${P}d$_3${S}$_4$},\
  }\href@noop {} {\bibfield  {journal} {\bibinfo  {journal} {Phys. Rev. B}\
  }\textbf {\bibinfo {volume} {107}},\ \bibinfo {pages} {014402} (\bibinfo
  {year} {2023})}\BibitemShut {NoStop}%
\bibitem [{\citenamefont {Hossain}\ \emph
  {et~al.}(2004{\natexlab{a}})\citenamefont {Hossain}, \citenamefont
  {Str\"assel}, \citenamefont {Geibel},\ and\ \citenamefont
  {Furrer}}]{Hossain2004VT}%
  \BibitemOpen
  \bibfield  {author} {\bibinfo {author} {\bibfnamefont {Z.}~\bibnamefont
  {Hossain}}, \bibinfo {author} {\bibfnamefont {T.}~\bibnamefont {Str\"assel}},
  \bibinfo {author} {\bibfnamefont {C.}~\bibnamefont {Geibel}},\ and\ \bibinfo
  {author} {\bibfnamefont {A.}~\bibnamefont {Furrer}},\ }\bibfield  {title}
  {\bibinfo {title} {{F}irst-order valence transition and barocaloric effect in
  {E}u{N}i$_2$({S}i$_{1-x}${G}e$_x$)$_2$},\ }\href@noop {} {\bibfield
  {journal} {\bibinfo  {journal} {Jour. Mag. Mag. Mat.}\ }\textbf {\bibinfo
  {volume} {272-276}},\ \bibinfo {pages} {2352} (\bibinfo {year}
  {2004}{\natexlab{a}})}\BibitemShut {NoStop}%
\bibitem [{\citenamefont {Ahmida}\ \emph {et~al.}(2020)\citenamefont {Ahmida},
  \citenamefont {Forthaus}, \citenamefont {Geibel}, \citenamefont {Hossain},
  \citenamefont {Hearne}, \citenamefont {Kastil}, \citenamefont {Prchal},
  \citenamefont {Sechovsky},\ and\ \citenamefont
  {Abd-Elmeguid}}]{Ahmida2020VF}%
  \BibitemOpen
  \bibfield  {author} {\bibinfo {author} {\bibfnamefont {M.~A.}\ \bibnamefont
  {Ahmida}}, \bibinfo {author} {\bibfnamefont {M.~K.}\ \bibnamefont
  {Forthaus}}, \bibinfo {author} {\bibfnamefont {C.}~\bibnamefont {Geibel}},
  \bibinfo {author} {\bibfnamefont {Z.}~\bibnamefont {Hossain}}, \bibinfo
  {author} {\bibfnamefont {G.~R.}\ \bibnamefont {Hearne}}, \bibinfo {author}
  {\bibfnamefont {J.}~\bibnamefont {Kastil}}, \bibinfo {author} {\bibfnamefont
  {J.}~\bibnamefont {Prchal}}, \bibinfo {author} {\bibfnamefont
  {V.}~\bibnamefont {Sechovsky}},\ and\ \bibinfo {author} {\bibfnamefont
  {M.~M.}\ \bibnamefont {Abd-Elmeguid}},\ }\bibfield  {title} {\bibinfo {title}
  {{C}harge fluctuations across the pressure-induced quantum phase transition
  in {E}u{C}u$_2$({S}i$_{1-x}${G}e$_x$)$_2$},\ }\href@noop {} {\bibfield
  {journal} {\bibinfo  {journal} {Phys. Rev. B}\ }\textbf {\bibinfo {volume}
  {101}},\ \bibinfo {pages} {205127} (\bibinfo {year} {2020})}\BibitemShut
  {NoStop}%
\bibitem [{\citenamefont {Baenitz}\ \emph {et~al.}(2006)\citenamefont
  {Baenitz}, \citenamefont {Gippius}, \citenamefont {Rajarajan}, \citenamefont
  {Morozova}, \citenamefont {Hossain}, \citenamefont {Geibel},\ and\
  \citenamefont {Steglich}}]{Baenitz2006NMR}%
  \BibitemOpen
  \bibfield  {author} {\bibinfo {author} {\bibfnamefont {M.}~\bibnamefont
  {Baenitz}}, \bibinfo {author} {\bibfnamefont {A.~A.}\ \bibnamefont
  {Gippius}}, \bibinfo {author} {\bibfnamefont {A.~K.}\ \bibnamefont
  {Rajarajan}}, \bibinfo {author} {\bibfnamefont {E.~N.}\ \bibnamefont
  {Morozova}}, \bibinfo {author} {\bibfnamefont {Z.}~\bibnamefont {Hossain}},
  \bibinfo {author} {\bibfnamefont {C.}~\bibnamefont {Geibel}},\ and\ \bibinfo
  {author} {\bibfnamefont {F.}~\bibnamefont {Steglich}},\ }\bibfield  {title}
  {\bibinfo {title} {{C}rossover from divalent to valence fluctuating state of
  {E}u in {E}u{C}u$_2$({S}i$_{1-x}${G}e$_x$)$_2$ probed by $^{63,
  65}${C}u-{NMR}},\ }\href@noop {} {\bibfield  {journal} {\bibinfo  {journal}
  {Physica B}\ }\textbf {\bibinfo {volume} {378-380}},\ \bibinfo {pages} {683}
  (\bibinfo {year} {2006})}\BibitemShut {NoStop}%
\bibitem [{\citenamefont {Dionicio}\ \emph {et~al.}(2006)\citenamefont
  {Dionicio}, \citenamefont {Wilhelm}, \citenamefont {Hossain},\ and\
  \citenamefont {Geibel}}]{Dionicio2006VT}%
  \BibitemOpen
  \bibfield  {author} {\bibinfo {author} {\bibfnamefont {M.}~\bibnamefont
  {Dionicio}}, \bibinfo {author} {\bibfnamefont {H.}~\bibnamefont {Wilhelm}},
  \bibinfo {author} {\bibfnamefont {Z.}~\bibnamefont {Hossain}},\ and\ \bibinfo
  {author} {\bibfnamefont {C.}~\bibnamefont {Geibel}},\ }\bibfield  {title}
  {\bibinfo {title} {Temperatur- and pressure-induced valence transition in
  {E}u{C}o$_2${G}e$_2$},\ }\href@noop {} {\bibfield  {journal} {\bibinfo
  {journal} {Physica B}\ }\textbf {\bibinfo {volume} {378-380}},\ \bibinfo
  {pages} {724} (\bibinfo {year} {2006})}\BibitemShut {NoStop}%
\bibitem [{\citenamefont {Hossain}\ \emph
  {et~al.}(2004{\natexlab{b}})\citenamefont {Hossain}, \citenamefont {Geibel},
  \citenamefont {Senthilkumaran}, \citenamefont {Deppe}, \citenamefont
  {Baenitz}, \citenamefont {Schiller},\ and\ \citenamefont
  {Molodtsov}}]{Hossain2004AFM}%
  \BibitemOpen
  \bibfield  {author} {\bibinfo {author} {\bibfnamefont {Z.}~\bibnamefont
  {Hossain}}, \bibinfo {author} {\bibfnamefont {C.}~\bibnamefont {Geibel}},
  \bibinfo {author} {\bibfnamefont {N.}~\bibnamefont {Senthilkumaran}},
  \bibinfo {author} {\bibfnamefont {M.}~\bibnamefont {Deppe}}, \bibinfo
  {author} {\bibfnamefont {M.}~\bibnamefont {Baenitz}}, \bibinfo {author}
  {\bibfnamefont {F.}~\bibnamefont {Schiller}},\ and\ \bibinfo {author}
  {\bibfnamefont {S.~L.}\ \bibnamefont {Molodtsov}},\ }\bibfield  {title}
  {\bibinfo {title} {{A}ntiferromagnetism, valence fluctuation, and
  heavy-fermion behavior in {E}u{C}u$_2$({S}i$_{1-x}${G}e$_x$)$_2$},\
  }\href@noop {} {\bibfield  {journal} {\bibinfo  {journal} {Phys. Rev. B}\
  }\textbf {\bibinfo {volume} {69}},\ \bibinfo {pages} {014422} (\bibinfo
  {year} {2004}{\natexlab{b}})}\BibitemShut {NoStop}%
\bibitem [{\citenamefont {Muthu}\ \emph {et~al.}(2016)\citenamefont {Muthu},
  \citenamefont {Braithwaite}, \citenamefont {Salce}, \citenamefont {Nakamura},
  \citenamefont {Hedo}, \citenamefont {Nakama},\ and\ \citenamefont
  {Onuki}}]{Muthu2016phase}%
  \BibitemOpen
  \bibfield  {author} {\bibinfo {author} {\bibfnamefont {S.~E.}\ \bibnamefont
  {Muthu}}, \bibinfo {author} {\bibfnamefont {D.}~\bibnamefont {Braithwaite}},
  \bibinfo {author} {\bibfnamefont {B.}~\bibnamefont {Salce}}, \bibinfo
  {author} {\bibfnamefont {A.}~\bibnamefont {Nakamura}}, \bibinfo {author}
  {\bibfnamefont {M.}~\bibnamefont {Hedo}}, \bibinfo {author} {\bibfnamefont
  {T.}~\bibnamefont {Nakama}},\ and\ \bibinfo {author} {\bibfnamefont
  {Y.}~\bibnamefont {Onuki}},\ }\bibfield  {title} {\bibinfo {title}
  {{C}alorimetry {S}tudy of the {P}hase {D}iagram of {E}u{N}i$_2${G}e$_2$},\
  }\href@noop {} {\bibfield  {journal} {\bibinfo  {journal} {J. Phys. Soc.
  Jpn.}\ }\textbf {\bibinfo {volume} {85}},\ \bibinfo {pages} {094603}
  (\bibinfo {year} {2016})}\BibitemShut {NoStop}%
\bibitem [{\citenamefont {Nakamura}\ \emph {et~al.}(2015)\citenamefont
  {Nakamura}, \citenamefont {Okazaki}, \citenamefont {Nakashima}, \citenamefont
  {Amako}, \citenamefont {Matsubayashi}, \citenamefont {Uwatoko}, \citenamefont
  {Kayama}, \citenamefont {T.~Kagayama}, \citenamefont {Uejo}, \citenamefont
  {Akamine}, \citenamefont {Hedo}, \citenamefont {Nakama},\ and\ \citenamefont
  {Onuki}}]{Nakamura2015trans}%
  \BibitemOpen
  \bibfield  {author} {\bibinfo {author} {\bibfnamefont {A.}~\bibnamefont
  {Nakamura}}, \bibinfo {author} {\bibfnamefont {T.}~\bibnamefont {Okazaki}},
  \bibinfo {author} {\bibfnamefont {M.}~\bibnamefont {Nakashima}}, \bibinfo
  {author} {\bibfnamefont {Y.}~\bibnamefont {Amako}}, \bibinfo {author}
  {\bibfnamefont {K.}~\bibnamefont {Matsubayashi}}, \bibinfo {author}
  {\bibfnamefont {Y.}~\bibnamefont {Uwatoko}}, \bibinfo {author} {\bibfnamefont
  {S.}~\bibnamefont {Kayama}}, \bibinfo {author} {\bibfnamefont {a.~K.~S.}\
  \bibnamefont {T.~Kagayama}}, \bibinfo {author} {\bibfnamefont
  {T.}~\bibnamefont {Uejo}}, \bibinfo {author} {\bibfnamefont {H.}~\bibnamefont
  {Akamine}}, \bibinfo {author} {\bibfnamefont {M.}~\bibnamefont {Hedo}},
  \bibinfo {author} {\bibfnamefont {T.}~\bibnamefont {Nakama}},\ and\ \bibinfo
  {author} {\bibfnamefont {Y.}~\bibnamefont {Onuki}},\ }\bibfield  {title}
  {\bibinfo {title} {{P}ressure-{I}nduced {V}alence {T}ransition and {H}eavy
  {F}ermion {S}tate in {E}u{N}i$_3${G}e$_5$ and {E}u{R}h{S}i$_3$},\ }\href@noop
  {} {\bibfield  {journal} {\bibinfo  {journal} {J. Phys. Soc. Jpn.}\ }\textbf
  {\bibinfo {volume} {84}},\ \bibinfo {pages} {053701} (\bibinfo {year}
  {2015})}\BibitemShut {NoStop}%
\bibitem [{\citenamefont {Gouchi}\ \emph {et~al.}(2020)\citenamefont {Gouchi},
  \citenamefont {Miyake}, \citenamefont {Iha}, \citenamefont {Hedo},
  \citenamefont {Nakama}, \citenamefont {Onuki},\ and\ \citenamefont
  {Uwatoko}}]{Gouchi2020AFM}%
  \BibitemOpen
  \bibfield  {author} {\bibinfo {author} {\bibfnamefont {J.}~\bibnamefont
  {Gouchi}}, \bibinfo {author} {\bibfnamefont {K.}~\bibnamefont {Miyake}},
  \bibinfo {author} {\bibfnamefont {W.}~\bibnamefont {Iha}}, \bibinfo {author}
  {\bibfnamefont {M.}~\bibnamefont {Hedo}}, \bibinfo {author} {\bibfnamefont
  {T.}~\bibnamefont {Nakama}}, \bibinfo {author} {\bibfnamefont
  {Y.}~\bibnamefont {Onuki}},\ and\ \bibinfo {author} {\bibfnamefont
  {Y.}~\bibnamefont {Uwatoko}},\ }\bibfield  {title} {\bibinfo {title}
  {{Q}uantum {C}riticality of {V}alence {T}ransition for the {U}nique
  {E}lectronic {S}tate of {A}ntiferromagnetic {C}ompound
  {E}u{C}u$_2${G}e$_2$},\ }\href@noop {} {\bibfield  {journal} {\bibinfo
  {journal} {J. Phys. Soc. Jpn.}\ }\textbf {\bibinfo {volume} {89}},\ \bibinfo
  {pages} {053703} (\bibinfo {year} {2020})}\BibitemShut {NoStop}%
\bibitem [{\citenamefont {Fukuda}\ \emph {et~al.}(2003)\citenamefont {Fukuda},
  \citenamefont {Nakanuma}, \citenamefont {Sakurai}, \citenamefont {Mitsuda},
  \citenamefont {Isikawa}, \citenamefont {Ishikawa}, \citenamefont {Goto},\
  and\ \citenamefont {Yamamoto}}]{Fukuda2003}%
  \BibitemOpen
  \bibfield  {author} {\bibinfo {author} {\bibfnamefont {S.}~\bibnamefont
  {Fukuda}}, \bibinfo {author} {\bibfnamefont {Y.}~\bibnamefont {Nakanuma}},
  \bibinfo {author} {\bibfnamefont {J.}~\bibnamefont {Sakurai}}, \bibinfo
  {author} {\bibfnamefont {A.}~\bibnamefont {Mitsuda}}, \bibinfo {author}
  {\bibfnamefont {Y.}~\bibnamefont {Isikawa}}, \bibinfo {author} {\bibfnamefont
  {F.}~\bibnamefont {Ishikawa}}, \bibinfo {author} {\bibfnamefont
  {T.}~\bibnamefont {Goto}},\ and\ \bibinfo {author} {\bibfnamefont
  {T.}~\bibnamefont {Yamamoto}},\ }\bibfield  {title} {\bibinfo {title}
  {{A}pplication of {D}oniach {D}iagram on {V}alence {T}ransition in
  {E}u{C}u$_2$({S}i$_{1-x}${G}e$_x$)$_2$},\ }\href@noop {} {\bibfield
  {journal} {\bibinfo  {journal} {J. Phys. Soc. Jpn.}\ }\textbf {\bibinfo
  {volume} {72}},\ \bibinfo {pages} {3189} (\bibinfo {year}
  {2003})}\BibitemShut {NoStop}%
\bibitem [{\citenamefont {Onuki}\ \emph {et~al.}(2017)\citenamefont {Onuki},
  \citenamefont {Nakamura}, \citenamefont {Aoki}, \citenamefont {Tekeuchoi},
  \citenamefont {Nakahima}, \citenamefont {Amako}, \citenamefont {Harima},
  \citenamefont {Matsubayashi}, \citenamefont {Uwatoko}, \citenamefont
  {Kayama}, \citenamefont {Kagayama}, \citenamefont {Shimizu}, \citenamefont
  {Esakki~Muthu}, \citenamefont {Braithwaite}, \citenamefont {Salce},
  \citenamefont {Shiba}, \citenamefont {Yara}, \citenamefont {Ashitomi},
  \citenamefont {Tomori}, \citenamefont {Hedo},\ and\ \citenamefont
  {Nakama}}]{Onuki2017review}%
  \BibitemOpen
  \bibfield  {author} {\bibinfo {author} {\bibfnamefont {A.}~\bibnamefont
  {Onuki}}, \bibinfo {author} {\bibfnamefont {F.}~\bibnamefont {Nakamura}},
  \bibinfo {author} {\bibfnamefont {T.}~\bibnamefont {Aoki}}, \bibinfo {author}
  {\bibfnamefont {T.}~\bibnamefont {Tekeuchoi}}, \bibinfo {author}
  {\bibfnamefont {M.}~\bibnamefont {Nakahima}}, \bibinfo {author}
  {\bibfnamefont {Y.}~\bibnamefont {Amako}}, \bibinfo {author} {\bibfnamefont
  {K.}~\bibnamefont {Harima}}, \bibinfo {author} {\bibfnamefont
  {K.}~\bibnamefont {Matsubayashi}}, \bibinfo {author} {\bibfnamefont
  {Y.}~\bibnamefont {Uwatoko}}, \bibinfo {author} {\bibfnamefont
  {S.}~\bibnamefont {Kayama}}, \bibinfo {author} {\bibfnamefont
  {T.}~\bibnamefont {Kagayama}}, \bibinfo {author} {\bibfnamefont
  {K.}~\bibnamefont {Shimizu}}, \bibinfo {author} {\bibfnamefont
  {S.}~\bibnamefont {Esakki~Muthu}}, \bibinfo {author} {\bibfnamefont
  {D.}~\bibnamefont {Braithwaite}}, \bibinfo {author} {\bibfnamefont
  {B.}~\bibnamefont {Salce}}, \bibinfo {author} {\bibfnamefont
  {H.}~\bibnamefont {Shiba}}, \bibinfo {author} {\bibfnamefont
  {T.}~\bibnamefont {Yara}}, \bibinfo {author} {\bibfnamefont {Y.}~\bibnamefont
  {Ashitomi}}, \bibinfo {author} {\bibfnamefont {K.}~\bibnamefont {Tomori}},
  \bibinfo {author} {\bibfnamefont {M.}~\bibnamefont {Hedo}},\ and\ \bibinfo
  {author} {\bibfnamefont {T.}~\bibnamefont {Nakama}},\ }\bibfield  {title}
  {\bibinfo {title} {Divalent, trivalent and heavy fermion states in {E}u
  compounds},\ }\href@noop {} {\bibfield  {journal} {\bibinfo  {journal}
  {Philosophical Magazine}\ }\textbf {\bibinfo {volume} {97}},\ \bibinfo
  {pages} {3399} (\bibinfo {year} {2017})}\BibitemShut {NoStop}%
\bibitem [{\citenamefont {Wada}\ \emph
  {et~al.}(1999{\natexlab{a}})\citenamefont {Wada}, \citenamefont {Sakata},
  \citenamefont {Nakamura}, \citenamefont {Mitsuda}, \citenamefont {Shiga},
  \citenamefont {Ikeda},\ and\ \citenamefont {Bando}}]{Wada1999phase}%
  \BibitemOpen
  \bibfield  {author} {\bibinfo {author} {\bibfnamefont {H.}~\bibnamefont
  {Wada}}, \bibinfo {author} {\bibfnamefont {T.}~\bibnamefont {Sakata}},
  \bibinfo {author} {\bibfnamefont {A.}~\bibnamefont {Nakamura}}, \bibinfo
  {author} {\bibfnamefont {A.}~\bibnamefont {Mitsuda}}, \bibinfo {author}
  {\bibfnamefont {M.}~\bibnamefont {Shiga}}, \bibinfo {author} {\bibfnamefont
  {Y.}~\bibnamefont {Ikeda}},\ and\ \bibinfo {author} {\bibfnamefont
  {Y.}~\bibnamefont {Bando}},\ }\bibfield  {title} {\bibinfo {title} {Thermal
  expansion and electric resistivity of
  {E}u{N}i$_2$({S}i$_{1-x}${G}e$_x$)$_2$},\ }\href@noop {} {\bibfield
  {journal} {\bibinfo  {journal} {J. Phys. Soc. Jpn.}\ }\textbf {\bibinfo
  {volume} {68}},\ \bibinfo {pages} {950} (\bibinfo {year}
  {1999}{\natexlab{a}})}\BibitemShut {NoStop}%
\bibitem [{\citenamefont {Onuki}\ \emph {et~al.}(2020)\citenamefont {Onuki},
  \citenamefont {Hedo},\ and\ \citenamefont {Honda}}]{Onuki2020phase}%
  \BibitemOpen
  \bibfield  {author} {\bibinfo {author} {\bibfnamefont {Y.}~\bibnamefont
  {Onuki}}, \bibinfo {author} {\bibfnamefont {M.}~\bibnamefont {Hedo}},\ and\
  \bibinfo {author} {\bibfnamefont {F.}~\bibnamefont {Honda}},\ }\bibfield
  {title} {\bibinfo {title} {Unique {E}lectronic {S}tates of {E}u-based
  {C}ompounds},\ }\href {https://doi.org/10.7566/JPSJ.89.102001} {\bibfield
  {journal} {\bibinfo  {journal} {J. Phys. Soc. Jpn.}\ }\textbf {\bibinfo
  {volume} {89}},\ \bibinfo {pages} {102001} (\bibinfo {year}
  {2020})}\BibitemShut {NoStop}%
\bibitem [{\citenamefont {Cho}\ \emph {et~al.}(2002)\citenamefont {Cho},
  \citenamefont {Rhyee},\ and\ \citenamefont {Ri}}]{Cho2002poly}%
  \BibitemOpen
  \bibfield  {author} {\bibinfo {author} {\bibfnamefont {B.~K.}\ \bibnamefont
  {Cho}}, \bibinfo {author} {\bibfnamefont {J.~S.}\ \bibnamefont {Rhyee}},\
  and\ \bibinfo {author} {\bibfnamefont {H.~C.}\ \bibnamefont {Ri}},\
  }\bibfield  {title} {\bibinfo {title} {Antiferromagnetic order and valence
  fluctuation in {E}u{P}d$_2$({G}e$_{1-x}${S}i$_x$)$_2$},\ }\href@noop {}
  {\bibfield  {journal} {\bibinfo  {journal} {J. Phys. Soc. Jpn.}\ }\textbf
  {\bibinfo {volume} {71}},\ \bibinfo {pages} {252} (\bibinfo {year}
  {2002})}\BibitemShut {NoStop}%
\bibitem [{\citenamefont {Seiro}\ and\ \citenamefont
  {Geibel}(2011)}]{Seiro2011EuIr2Si2}%
  \BibitemOpen
  \bibfield  {author} {\bibinfo {author} {\bibfnamefont {S.}~\bibnamefont
  {Seiro}}\ and\ \bibinfo {author} {\bibfnamefont {C.}~\bibnamefont {Geibel}},\
  }\bibfield  {title} {\bibinfo {title} {From stable divalent to
  valencefluctuating behaviour in {E}u({R}h$_{1-x}${I}r$_x$)$_2${S}i$_2$ single
  crystals},\ }\href@noop {} {\bibfield  {journal} {\bibinfo  {journal} {J.
  Phys.: Condens. Matter}\ }\textbf {\bibinfo {volume} {23}},\ \bibinfo {pages}
  {375601} (\bibinfo {year} {2011})}\BibitemShut {NoStop}%
\bibitem [{\citenamefont {Batlogg}\ \emph {et~al.}(1982)\citenamefont
  {Batlogg}, \citenamefont {Jayaraman}, \citenamefont {Murgai}, \citenamefont
  {Gupta}, \citenamefont {Parks},\ and\ \citenamefont
  {Croft}}]{Batlogg1982EuPd2Si2}%
  \BibitemOpen
  \bibfield  {author} {\bibinfo {author} {\bibfnamefont {B.}~\bibnamefont
  {Batlogg}}, \bibinfo {author} {\bibfnamefont {A.}~\bibnamefont {Jayaraman}},
  \bibinfo {author} {\bibfnamefont {V.}~\bibnamefont {Murgai}}, \bibinfo
  {author} {\bibfnamefont {L.~C.}\ \bibnamefont {Gupta}}, \bibinfo {author}
  {\bibfnamefont {R.~D.}\ \bibnamefont {Parks}},\ and\ \bibinfo {author}
  {\bibfnamefont {M.}~\bibnamefont {Croft}},\ }\bibfield  {title} {\bibinfo
  {title} {Pressure - temperature studies and the p{T}-diagram of
  {E}u{P}d$_2${S}i$_2$},\ }\href@noop {} {\bibfield  {journal} {\bibinfo
  {journal} {in $\textit{Valence Instabilities}$ ed. P. Wachter, and H.
  Boppart}\ ,\ \bibinfo {pages} {229}} (\bibinfo {year} {1982})}\BibitemShut
  {NoStop}%
\bibitem [{\citenamefont {Mimura}\ \emph {et~al.}(2004)\citenamefont {Mimura},
  \citenamefont {Taguchi}, \citenamefont {Sakurai},\ and\ \citenamefont
  {Aita}}]{Mimura2004EuPd2Si2}%
  \BibitemOpen
  \bibfield  {author} {\bibinfo {author} {\bibfnamefont {K.}~\bibnamefont
  {Mimura}}, \bibinfo {author} {\bibfnamefont {S.~M.~A.}\ \bibnamefont
  {Taguchi}, \bibfnamefont {Y.~Fukuda}}, \bibinfo {author} {\bibfnamefont
  {K.}~\bibnamefont {Sakurai}, \bibfnamefont {J.~Ichikawa}},\ and\ \bibinfo
  {author} {\bibfnamefont {O.}~\bibnamefont {Aita}},\ }\bibfield  {title}
  {\bibinfo {title} {Bulk-sensitive high-resolution photoemission study of a
  temperature-induced valence transition system in {E}u{P}d$_2${S}i$_2$},\
  }\href@noop {} {\bibfield  {journal} {\bibinfo  {journal} {Electron
  Spectroscopy and Related Phenomena}\ }\textbf {\bibinfo {volume} {137-140}},\
  \bibinfo {pages} {529} (\bibinfo {year} {2004})}\BibitemShut {NoStop}%
\bibitem [{\citenamefont {Felner}\ and\ \citenamefont
  {Nowik}(1986)}]{Felner1986YbInCu4}%
  \BibitemOpen
  \bibfield  {author} {\bibinfo {author} {\bibfnamefont {I.}~\bibnamefont
  {Felner}}\ and\ \bibinfo {author} {\bibfnamefont {I.}~\bibnamefont {Nowik}},\
  }\bibfield  {title} {\bibinfo {title} {First-order valence phase transition
  in cubic {Y}b{I}n{C}u$_4$},\ }\href@noop {} {\bibfield  {journal} {\bibinfo
  {journal} {Phys. Rev. B}\ }\textbf {\bibinfo {volume} {33}},\ \bibinfo
  {pages} {617} (\bibinfo {year} {1986})}\BibitemShut {NoStop}%
\bibitem [{\citenamefont {Sampathkumaran}\ \emph {et~al.}(1981)\citenamefont
  {Sampathkumaran}, \citenamefont {Vijayaraghavan}, \citenamefont
  {Gopalakrishnan}, \citenamefont {Pillay}, \citenamefont {Devare},
  \citenamefont {Gupta}, \citenamefont {Post},\ and\ \citenamefont
  {Parks}}]{Sampath1981EuPd2Si2}%
  \BibitemOpen
  \bibfield  {author} {\bibinfo {author} {\bibfnamefont {E.~V.}\ \bibnamefont
  {Sampathkumaran}}, \bibinfo {author} {\bibfnamefont {K.}~\bibnamefont
  {Vijayaraghavan}}, \bibinfo {author} {\bibfnamefont {V.}~\bibnamefont
  {Gopalakrishnan}}, \bibinfo {author} {\bibfnamefont {R.}~\bibnamefont
  {Pillay}}, \bibinfo {author} {\bibfnamefont {H.}~\bibnamefont {Devare}},
  \bibinfo {author} {\bibfnamefont {L.}~\bibnamefont {Gupta}}, \bibinfo
  {author} {\bibfnamefont {B.}~\bibnamefont {Post}},\ and\ \bibinfo {author}
  {\bibfnamefont {R.}~\bibnamefont {Parks}},\ }\bibfield  {title} {\bibinfo
  {title} {Valence transition of {E}u{P}d$_2${S}i$_2$},\ }\href@noop {}
  {\bibfield  {journal} {\bibinfo  {journal} {in $\textit{Valence fluctuations
  in solids}$ ed. M. L. Falicov, and W. Hanke, and M. B. Maple}\ ,\ \bibinfo
  {pages} {193}} (\bibinfo {year} {1981})}\BibitemShut {NoStop}%
\bibitem [{\citenamefont {Croft}\ \emph {et~al.}(1982)\citenamefont {Croft},
  \citenamefont {Hodges}, \citenamefont {Kemly}, \citenamefont {Krishnan},
  \citenamefont {Murgai},\ and\ \citenamefont {Gupta}}]{Croft1982config}%
  \BibitemOpen
  \bibfield  {author} {\bibinfo {author} {\bibfnamefont {M.}~\bibnamefont
  {Croft}}, \bibinfo {author} {\bibfnamefont {J.~A.}\ \bibnamefont {Hodges}},
  \bibinfo {author} {\bibfnamefont {E.}~\bibnamefont {Kemly}}, \bibinfo
  {author} {\bibfnamefont {A.}~\bibnamefont {Krishnan}}, \bibinfo {author}
  {\bibfnamefont {V.}~\bibnamefont {Murgai}},\ and\ \bibinfo {author}
  {\bibfnamefont {L.~C.}\ \bibnamefont {Gupta}},\ }\bibfield  {title} {\bibinfo
  {title} {Cooperative configuration change in {E}u{P}d$_2${S}i$_2$},\
  }\href@noop {} {\bibfield  {journal} {\bibinfo  {journal} {Phys. Rev. Lett.}\
  }\textbf {\bibinfo {volume} {48}},\ \bibinfo {pages} {826} (\bibinfo {year}
  {1982})}\BibitemShut {NoStop}%
\bibitem [{\citenamefont {Kliemt}\ \emph {et~al.}(2022)\citenamefont {Kliemt},
  \citenamefont {Peters}, \citenamefont {Reiser}, \citenamefont {Ocker},
  \citenamefont {Walther}, \citenamefont {Tran}, \citenamefont {Cho},
  \citenamefont {Merz}, \citenamefont {Haghighirad}, \citenamefont {Hezel},
  \citenamefont {Ritter},\ and\ \citenamefont {Krellner}}]{Kliemt2022cryst}%
  \BibitemOpen
  \bibfield  {author} {\bibinfo {author} {\bibfnamefont {K.}~\bibnamefont
  {Kliemt}}, \bibinfo {author} {\bibfnamefont {M.}~\bibnamefont {Peters}},
  \bibinfo {author} {\bibfnamefont {I.}~\bibnamefont {Reiser}}, \bibinfo
  {author} {\bibfnamefont {M.}~\bibnamefont {Ocker}}, \bibinfo {author}
  {\bibfnamefont {F.}~\bibnamefont {Walther}}, \bibinfo {author} {\bibfnamefont
  {D.-M.}\ \bibnamefont {Tran}}, \bibinfo {author} {\bibfnamefont
  {E.}~\bibnamefont {Cho}}, \bibinfo {author} {\bibfnamefont {M.}~\bibnamefont
  {Merz}}, \bibinfo {author} {\bibfnamefont {A.}~\bibnamefont {Haghighirad}},
  \bibinfo {author} {\bibfnamefont {D.}~\bibnamefont {Hezel}}, \bibinfo
  {author} {\bibfnamefont {F.}~\bibnamefont {Ritter}},\ and\ \bibinfo {author}
  {\bibfnamefont {C.}~\bibnamefont {Krellner}},\ }\bibfield  {title} {\bibinfo
  {title} {Strong influence of the {P}d-{S}i ratio on the valence transition in
  {E}u{P}d$_2${S}i$_2$ single crystals},\ }\href@noop {} {\bibfield  {journal}
  {\bibinfo  {journal} {Cryst. Growth Des.}\ }\textbf {\bibinfo {volume}
  {22}},\ \bibinfo {pages} {5399} (\bibinfo {year} {2022})}\BibitemShut
  {NoStop}%
\bibitem [{\citenamefont {Peters}\ \emph {et~al.}()\citenamefont {Peters},
  \citenamefont {Kliemt}, \citenamefont {Ocker}, \citenamefont {Wolf},
  \citenamefont {Puphal}, \citenamefont {Le~Tacon}, \citenamefont {Merz},
  \citenamefont {Lang},\ and\ \citenamefont {Krellner}}]{Peters2022cryst}%
  \BibitemOpen
  \bibfield  {author} {\bibinfo {author} {\bibfnamefont {M.}~\bibnamefont
  {Peters}}, \bibinfo {author} {\bibfnamefont {K.}~\bibnamefont {Kliemt}},
  \bibinfo {author} {\bibfnamefont {M.}~\bibnamefont {Ocker}}, \bibinfo
  {author} {\bibfnamefont {B.}~\bibnamefont {Wolf}}, \bibinfo {author}
  {\bibfnamefont {P.}~\bibnamefont {Puphal}}, \bibinfo {author} {\bibfnamefont
  {M.}~\bibnamefont {Le~Tacon}}, \bibinfo {author} {\bibfnamefont
  {M.}~\bibnamefont {Merz}}, \bibinfo {author} {\bibfnamefont {M.}~\bibnamefont
  {Lang}},\ and\ \bibinfo {author} {\bibfnamefont {C.}~\bibnamefont
  {Krellner}},\ }\bibfield  {title} {\bibinfo {title} {{F}rom valence
  fluctuations to long-range magnetic order in
  {E}u{P}d$_2$({S}i$_{1-x}${G}e$_x$)$_2$ single crystals},\ }\href@noop {} {\
  }\Eprint {https://arxiv.org/abs/http://arxiv.org/2303.07472v2, accepted for
  publication} {http://arxiv.org/2303.07472v2, accepted for publication}
  \BibitemShut {NoStop}%
\bibitem [{\citenamefont {Wolf}\ \emph
  {et~al.}(2022{\natexlab{a}})\citenamefont {Wolf}, \citenamefont {Spathelf},
  \citenamefont {Zimmermann}, \citenamefont {Lundbeck}, \citenamefont {Peters},
  \citenamefont {Kliemt}, \citenamefont {Krellner},\ and\ \citenamefont
  {Lang}}]{Wolf2022SCES}%
  \BibitemOpen
  \bibfield  {author} {\bibinfo {author} {\bibfnamefont {B.}~\bibnamefont
  {Wolf}}, \bibinfo {author} {\bibfnamefont {F.}~\bibnamefont {Spathelf}},
  \bibinfo {author} {\bibfnamefont {J.}~\bibnamefont {Zimmermann}}, \bibinfo
  {author} {\bibfnamefont {T.}~\bibnamefont {Lundbeck}}, \bibinfo {author}
  {\bibfnamefont {M.}~\bibnamefont {Peters}}, \bibinfo {author} {\bibfnamefont
  {K.}~\bibnamefont {Kliemt}}, \bibinfo {author} {\bibfnamefont
  {C.}~\bibnamefont {Krellner}},\ and\ \bibinfo {author} {\bibfnamefont
  {M.}~\bibnamefont {Lang}},\ }\href@noop {} {\bibfield  {journal} {\bibinfo
  {journal} {SciPost}\ }\textbf {\bibinfo {volume} {202207-00023v2}} (\bibinfo
  {year} {2022}{\natexlab{a}})}\BibitemShut {NoStop}%
\bibitem [{\citenamefont {Segre}\ \emph {et~al.}(1982)\citenamefont {Segre},
  \citenamefont {Croft}, \citenamefont {Hodges}, \citenamefont {Murgai},
  \citenamefont {Gupta},\ and\ \citenamefont {Parks}}]{Segre1982mix}%
  \BibitemOpen
  \bibfield  {author} {\bibinfo {author} {\bibfnamefont {C.~U.}\ \bibnamefont
  {Segre}}, \bibinfo {author} {\bibfnamefont {M.}~\bibnamefont {Croft}},
  \bibinfo {author} {\bibfnamefont {J.~A.}\ \bibnamefont {Hodges}}, \bibinfo
  {author} {\bibfnamefont {V.}~\bibnamefont {Murgai}}, \bibinfo {author}
  {\bibfnamefont {L.~C.}\ \bibnamefont {Gupta}},\ and\ \bibinfo {author}
  {\bibfnamefont {R.~D.}\ \bibnamefont {Parks}},\ }\bibfield  {title} {\bibinfo
  {title} {Valence instability in {E}u({P}d$_{1-x}${A}u$_x$)$_2${S}i$_2$. {T}he
  {G}lobal {P}hase {D}iagram},\ }\href@noop {} {\bibfield  {journal} {\bibinfo
  {journal} {Phys. Rev. Lett.}\ }\textbf {\bibinfo {volume} {49}},\ \bibinfo
  {pages} {1947} (\bibinfo {year} {1982})}\BibitemShut {NoStop}%
\bibitem [{\citenamefont {Ye}\ \emph {et~al.}()\citenamefont {Ye},
  \citenamefont {von Westarp}, \citenamefont {Souliou}, \citenamefont {Peters},
  \citenamefont {M\"oller}, \citenamefont {Kliemt}, \citenamefont {Merz},
  \citenamefont {Heid}, \citenamefont {Krellner},\ and\ \citenamefont
  {Tacon}}]{Ye2022Raman}%
  \BibitemOpen
  \bibfield  {author} {\bibinfo {author} {\bibfnamefont {M.}~\bibnamefont
  {Ye}}, \bibinfo {author} {\bibfnamefont {M.~J.~G.}\ \bibnamefont {von
  Westarp}}, \bibinfo {author} {\bibfnamefont {S.~M.}\ \bibnamefont {Souliou}},
  \bibinfo {author} {\bibfnamefont {M.}~\bibnamefont {Peters}}, \bibinfo
  {author} {\bibfnamefont {R.}~\bibnamefont {M\"oller}}, \bibinfo {author}
  {\bibfnamefont {K.}~\bibnamefont {Kliemt}}, \bibinfo {author} {\bibfnamefont
  {M.}~\bibnamefont {Merz}}, \bibinfo {author} {\bibfnamefont {R.}~\bibnamefont
  {Heid}}, \bibinfo {author} {\bibfnamefont {C.}~\bibnamefont {Krellner}},\
  and\ \bibinfo {author} {\bibfnamefont {M.~L.}\ \bibnamefont {Tacon}},\
  }\bibfield  {title} {\bibinfo {title} {{S}trong electron-phonon coupling and
  enhanced phonon {G}r\"uneisen parameters in valence-fluctuating metal
  {E}u{P}d$_2${S}i$_2$},\ }\Eprint
  {https://arxiv.org/abs/http://arxiv.org/2211.04229v1}
  {http://arxiv.org/2211.04229v1} \BibitemShut {NoStop}%
\bibitem [{\citenamefont {Wolf}\ \emph
  {et~al.}(2022{\natexlab{b}})\citenamefont {Wolf}, \citenamefont {Kaib},
  \citenamefont {Razpopov}, \citenamefont {Biswas}, \citenamefont {Riedl},
  \citenamefont {Winter}, \citenamefont {Valent\'{\i}}, \citenamefont {Saito},
  \citenamefont {Hartmann}, \citenamefont {Vinokurova}, \citenamefont {Doert},
  \citenamefont {Isaeva}, \citenamefont {Bastien}, \citenamefont {Wolter},
  \citenamefont {B\"uchner},\ and\ \citenamefont {Lang}}]{Wolf2022RuCl3}%
  \BibitemOpen
  \bibfield  {author} {\bibinfo {author} {\bibfnamefont {B.}~\bibnamefont
  {Wolf}}, \bibinfo {author} {\bibfnamefont {D.~A.~S.}\ \bibnamefont {Kaib}},
  \bibinfo {author} {\bibfnamefont {A.}~\bibnamefont {Razpopov}}, \bibinfo
  {author} {\bibfnamefont {S.}~\bibnamefont {Biswas}}, \bibinfo {author}
  {\bibfnamefont {K.}~\bibnamefont {Riedl}}, \bibinfo {author} {\bibfnamefont
  {S.~M.}\ \bibnamefont {Winter}}, \bibinfo {author} {\bibfnamefont
  {R.}~\bibnamefont {Valent\'{\i}}}, \bibinfo {author} {\bibfnamefont
  {Y.}~\bibnamefont {Saito}}, \bibinfo {author} {\bibfnamefont
  {S.}~\bibnamefont {Hartmann}}, \bibinfo {author} {\bibfnamefont
  {E.}~\bibnamefont {Vinokurova}}, \bibinfo {author} {\bibfnamefont
  {T.}~\bibnamefont {Doert}}, \bibinfo {author} {\bibfnamefont
  {A.}~\bibnamefont {Isaeva}}, \bibinfo {author} {\bibfnamefont
  {G.}~\bibnamefont {Bastien}}, \bibinfo {author} {\bibfnamefont {A.~U.~B.}\
  \bibnamefont {Wolter}}, \bibinfo {author} {\bibfnamefont {B.}~\bibnamefont
  {B\"uchner}},\ and\ \bibinfo {author} {\bibfnamefont {M.}~\bibnamefont
  {Lang}},\ }\bibfield  {title} {\bibinfo {title} {Combined experimental and
  theoretical study of hydrostatic ({H}e-gas) pressure effects in
  $\alpha$-{R}u{C}l$_3$},\ }\href@noop {} {\bibfield  {journal} {\bibinfo
  {journal} {Phys. Rev. B}\ }\textbf {\bibinfo {volume} {106}},\ \bibinfo
  {pages} {134432} (\bibinfo {year} {2022}{\natexlab{b}})}\BibitemShut
  {NoStop}%
\bibitem [{\citenamefont {Y.~Takikawa}(2010)}]{Takikawa2010}%
  \BibitemOpen
  \bibfield  {author} {\bibinfo {author} {\bibfnamefont {S.~N.}\ \bibnamefont
  {Y.~Takikawa}, \bibfnamefont {S.~Ebisu}},\ }\bibfield  {title} {\bibinfo
  {title} {Van {V}leck paramagnetism of the trivalent {E}u ions},\ }\href@noop
  {} {\bibfield  {journal} {\bibinfo  {journal} {Journal of Physics and
  Chemistry of Solids}\ }\textbf {\bibinfo {volume} {71}},\ \bibinfo {pages}
  {1592} (\bibinfo {year} {2010})}\BibitemShut {NoStop}%
\bibitem [{\citenamefont {Wada}\ \emph {et~al.}(2001)\citenamefont {Wada},
  \citenamefont {Gomi}, \citenamefont {Mitsuda},\ and\ \citenamefont
  {Shiga}}]{Wada2001mix}%
  \BibitemOpen
  \bibfield  {author} {\bibinfo {author} {\bibfnamefont {H.}~\bibnamefont
  {Wada}}, \bibinfo {author} {\bibfnamefont {H.}~\bibnamefont {Gomi}}, \bibinfo
  {author} {\bibfnamefont {A.}~\bibnamefont {Mitsuda}},\ and\ \bibinfo {author}
  {\bibfnamefont {M.}~\bibnamefont {Shiga}},\ }\bibfield  {title} {\bibinfo
  {title} {Specific heat anomaly due to valence transition in
  {E}u({P}d$_{1-x}${P}t$_x$)$_2${S}i$_2$},\ }\href@noop {} {\bibfield
  {journal} {\bibinfo  {journal} {Solid State Communications}\ }\textbf
  {\bibinfo {volume} {117}},\ \bibinfo {pages} {703} (\bibinfo {year}
  {2001})}\BibitemShut {NoStop}%
\bibitem [{Note1()}]{Note1}%
  \BibitemOpen
  \bibinfo {note} {Note that the susceptibility data of x = 0.058 (\#3) for $p$
  = 0.43\protect \,GPa and $p$ = 0.47\protect \,GPa were omitted in the inset
  of FIG.\protect \,3 for clarity.}\BibitemShut {Stop}%
\bibitem [{\citenamefont {Song}\ \emph {et~al.}(2023)\citenamefont {Song},
  \citenamefont {Schulz}, \citenamefont {Kliemt}, \citenamefont {Krellner},\
  and\ \citenamefont {Valenti}}]{Song2023DFT}%
  \BibitemOpen
  \bibfield  {author} {\bibinfo {author} {\bibfnamefont {Y.-J.}\ \bibnamefont
  {Song}}, \bibinfo {author} {\bibfnamefont {S.}~\bibnamefont {Schulz}},
  \bibinfo {author} {\bibfnamefont {K.}~\bibnamefont {Kliemt}}, \bibinfo
  {author} {\bibfnamefont {C.}~\bibnamefont {Krellner}},\ and\ \bibinfo
  {author} {\bibfnamefont {R.}~\bibnamefont {Valenti}},\ }\bibfield  {title}
  {\bibinfo {title} {Microscopic origin of the valence transition in tetragonal
  {E}u{P}d$_2${S}i$_2$},\ }\href@noop {} {\bibfield  {journal} {\bibinfo
  {journal} {Phys. Rev. B}\ }\textbf {\bibinfo {volume} {in press}} (\bibinfo
  {year} {2023})}\BibitemShut {NoStop}%
\bibitem [{\citenamefont {Wada}\ \emph
  {et~al.}(1999{\natexlab{b}})\citenamefont {Wada}, \citenamefont {Hundley},
  \citenamefont {Movshovich},\ and\ \citenamefont
  {Thompson}}]{Thomson1999pressure}%
  \BibitemOpen
  \bibfield  {author} {\bibinfo {author} {\bibfnamefont {H.}~\bibnamefont
  {Wada}}, \bibinfo {author} {\bibfnamefont {M.~F.}\ \bibnamefont {Hundley}},
  \bibinfo {author} {\bibfnamefont {R.}~\bibnamefont {Movshovich}},\ and\
  \bibinfo {author} {\bibfnamefont {J.~D.}\ \bibnamefont {Thompson}},\
  }\bibfield  {title} {\bibinfo {title} {Pressure effect on the valence
  transition of {E}u{N}i$_2$({S}i$_{1-x}${G}e$_x$)$_2$},\ }\href@noop {}
  {\bibfield  {journal} {\bibinfo  {journal} {Phys. Rev. B}\ }\textbf {\bibinfo
  {volume} {59}},\ \bibinfo {pages} {1141} (\bibinfo {year}
  {1999}{\natexlab{b}})}\BibitemShut {NoStop}%
\bibitem [{\citenamefont {Schmiester}\ \emph {et~al.}(1982)\citenamefont
  {Schmiester}, \citenamefont {Perscheid}, \citenamefont {Kaindl},\ and\
  \citenamefont {Zukrowsky}}]{Schmiester1982EuPd2Si2}%
  \BibitemOpen
  \bibfield  {author} {\bibinfo {author} {\bibfnamefont {B.~G.}\ \bibnamefont
  {Schmiester}}, \bibinfo {author} {\bibfnamefont {B.}~\bibnamefont
  {Perscheid}}, \bibinfo {author} {\bibfnamefont {G.}~\bibnamefont {Kaindl}},\
  and\ \bibinfo {author} {\bibfnamefont {J.}~\bibnamefont {Zukrowsky}},\
  }\bibfield  {title} {\bibinfo {title} {Effects of {P}ressure and
  {T}emperature on the mean valence of {E}u{P}d$_2${S}i$_2$},\ }\href@noop {}
  {\bibfield  {journal} {\bibinfo  {journal} {in $\textit{Valence
  Instabilities}$ ed. P. Wachter and H. Boppart}\ ,\ \bibinfo {pages} {219}}
  (\bibinfo {year} {1982})}\BibitemShut {NoStop}%
\bibitem [{\citenamefont {Anand}\ and\ \citenamefont
  {Johnston}(2015)}]{Anand2015AFM}%
  \BibitemOpen
  \bibfield  {author} {\bibinfo {author} {\bibfnamefont {V.~K.}\ \bibnamefont
  {Anand}}\ and\ \bibinfo {author} {\bibfnamefont {D.~C.}\ \bibnamefont
  {Johnston}},\ }\bibfield  {title} {\bibinfo {title} {Antiferromagnetism in
  {E}u{C}u$_2${A}s$_2$ and {E}u{C}u$_{1.82}${S}b$_2$ single crystals},\
  }\href@noop {} {\bibfield  {journal} {\bibinfo  {journal} {Phys. Rev. B}\
  }\textbf {\bibinfo {volume} {91}},\ \bibinfo {pages} {184403} (\bibinfo
  {year} {2015})}\BibitemShut {NoStop}%
\bibitem [{\citenamefont {Watanabe}\ and\ \citenamefont
  {Miyake}(2011)}]{Watanabe2011Theo}%
  \BibitemOpen
  \bibfield  {author} {\bibinfo {author} {\bibfnamefont {S.}~\bibnamefont
  {Watanabe}}\ and\ \bibinfo {author} {\bibfnamefont {K.}~\bibnamefont
  {Miyake}},\ }\bibfield  {title} {\bibinfo {title} {Roles of critical valence
  fluctuations in {C}e- and {Y}b-based heavy fermion metals},\ }\href
  {https://doi.org/http://dx.doi.org/10.1088/0953-8984/23/9/094217} {\bibfield
  {journal} {\bibinfo  {journal} {J. Phys.: Condens. Matter}\ }\textbf
  {\bibinfo {volume} {23}},\ \bibinfo {pages} {094217} (\bibinfo {year}
  {2011})}\BibitemShut {NoStop}%
\bibitem [{\citenamefont {Manna}\ \emph {et~al.}(2012)\citenamefont {Manna},
  \citenamefont {Wolf}, \citenamefont {de~Souza},\ and\ \citenamefont
  {Lang}}]{Manna2012druck}%
  \BibitemOpen
  \bibfield  {author} {\bibinfo {author} {\bibfnamefont {R.~S.}\ \bibnamefont
  {Manna}}, \bibinfo {author} {\bibfnamefont {B.}~\bibnamefont {Wolf}},
  \bibinfo {author} {\bibfnamefont {M.}~\bibnamefont {de~Souza}},\ and\
  \bibinfo {author} {\bibfnamefont {M.}~\bibnamefont {Lang}},\ }\bibfield
  {title} {\bibinfo {title} {High-resolution thermal expansion measurements
  under helium-gas pressure},\ }\href@noop {} {\bibfield  {journal} {\bibinfo
  {journal} {Rev. Sci. Instrum.}\ }\textbf {\bibinfo {volume} {83}},\ \bibinfo
  {pages} {08511} (\bibinfo {year} {2012})}\BibitemShut {NoStop}%
\bibitem [{\citenamefont {Agarmani}\ \emph {et~al.}(2022)\citenamefont
  {Agarmani}, \citenamefont {Hartmann}, \citenamefont {Zimmermann},
  \citenamefont {Gati}, \citenamefont {Delleske}, \citenamefont {Tutsch},
  \citenamefont {Wolf},\ and\ \citenamefont {Lang}}]{Agarmani2022druck}%
  \BibitemOpen
  \bibfield  {author} {\bibinfo {author} {\bibfnamefont {Y.}~\bibnamefont
  {Agarmani}}, \bibinfo {author} {\bibfnamefont {S.}~\bibnamefont {Hartmann}},
  \bibinfo {author} {\bibfnamefont {J.}~\bibnamefont {Zimmermann}}, \bibinfo
  {author} {\bibfnamefont {E.}~\bibnamefont {Gati}}, \bibinfo {author}
  {\bibfnamefont {C.}~\bibnamefont {Delleske}}, \bibinfo {author}
  {\bibfnamefont {U.}~\bibnamefont {Tutsch}}, \bibinfo {author} {\bibfnamefont
  {B.}~\bibnamefont {Wolf}},\ and\ \bibinfo {author} {\bibfnamefont
  {M.}~\bibnamefont {Lang}},\ }\bibfield  {title} {\bibinfo {title} {Advanced
  technique for measuring relative length changes under control of temperature
  and helium-gas pressure},\ }\href@noop {} {\bibfield  {journal} {\bibinfo
  {journal} {Rev. Sci. Instrum.}\ }\textbf {\bibinfo {volume} {93}},\ \bibinfo
  {pages} {113902} (\bibinfo {year} {2022})}\BibitemShut {NoStop}%
\bibitem [{\citenamefont {Fisher}(1962)}]{Fisher1962specHeat}%
  \BibitemOpen
  \bibfield  {author} {\bibinfo {author} {\bibfnamefont {M.~E.}\ \bibnamefont
  {Fisher}},\ }\bibfield  {title} {\bibinfo {title} {{R}elation between the
  {S}pecific {H}eat and {S}usceptibility of an {A}ntiferromagnet},\ }\href
  {https://doi.org/10.1080/14786436208213705} {\bibfield  {journal} {\bibinfo
  {journal} {{P}hilosophical {M}agazine}\ }\textbf {\bibinfo {volume} {7}},\
  \bibinfo {pages} {1731} (\bibinfo {year} {1962})}\BibitemShut {NoStop}%
\bibitem [{\citenamefont {Sales}\ and\ \citenamefont
  {Wohlleben}(1975)}]{Wohlleben1975ICF}%
  \BibitemOpen
  \bibfield  {author} {\bibinfo {author} {\bibfnamefont {B.~C.}\ \bibnamefont
  {Sales}}\ and\ \bibinfo {author} {\bibfnamefont {D.~K.}\ \bibnamefont
  {Wohlleben}},\ }\bibfield  {title} {\bibinfo {title} {Susceptibility of
  {I}nterconfiguration-{F}luctuation {C}ompounds},\ }\href@noop {} {\bibfield
  {journal} {\bibinfo  {journal} {Phys. Rev. Lett.}\ }\textbf {\bibinfo
  {volume} {35}},\ \bibinfo {pages} {1240} (\bibinfo {year}
  {1975})}\BibitemShut {NoStop}%
\end{thebibliography}%

\end{document}